\documentclass[%
 aps, prr,
 amsmath,amssymb,
 reprint,
 longbibliography
]{revtex4-1}

\usepackage{dcolumn}
\usepackage{bm}
\usepackage[dvipsnames]{xcolor}
\usepackage{soul}
\usepackage{amssymb,mathrsfs}
\usepackage{mathtools}
\usepackage[normalem]{ulem} 
\usepackage{physics}
\usepackage{comment}
\usepackage[dvipsnames]{xcolor}
\definecolor{refkey}{gray}{.75}
\definecolor{labelkey}{gray}{.75}
\usepackage{graphicx}

\usepackage[utf8]{inputenc}
\usepackage[T1]{fontenc}
\usepackage{mathptmx}
\usepackage{etoolbox}
\usepackage{placeins}

\newcommand{\revisiondiff}{0}
\newcommand{\revisionnew}{1}

\newcommand{\revisiontype}{\revisionnew}
\newcommand{\revision}[2]{%
\if\revisiontype\revisiondiff%
     {\color{red}\st{#1}}{\color{ForestGreen}#2}%
\else%
     \if\revisiontype\revisionnew%
         #2%
     \else%
         #1%
     \fi%
\fi}%

\makeatletter
\def\@email#1#2{%
 \endgroup
 \patchcmd{\titleblock@produce}
  {\frontmatter@RRAPformat}
  {\frontmatter@RRAPformat{\produce@RRAP{*#1\href{mailto:#2}{#2}}}\frontmatter@RRAPformat}
  {}{}
}%
\makeatother
\newcommand{\pref}[1]{(\ref{#1})}
\newcommand{\epref}[1]{Eq.~(\ref{#1})}
\newcommand{\figref}[1]{Fig.~\ref{#1}}

\newcommand{\appref}[1]{Appendix~\ref{#1}}

\newcommand{\ci}{{\mathrm i}}
\newcommand{\iu}{\mathrm i}
\newcommand{\rk}[1]{{\color{Plum}\textbf{[RK:} #1\textbf{]}}}
\definecolor{armygreen}{rgb}{0.29, 0.33, 0.13}

\newcommand{\ie}{\textit{i.e.}}
\newcommand{\ea}{\textit{et al.}}
\newcommand{\ai}{\textit{ab initio}}
\newcommand{\eg}{\textit{e.g.}}

\newcommand{\half}{\frac 12}
\newcommand{\EF}{{E_{\mathrm F}}}
\newcommand{\thetat}{\theta_{\mathrm{T}}}
\newcommand{\s}[1]{\hat\sigma_{#1}}
\newcommand{\acom}[2]{\qty{ #1 ,\, #2 }}
\newcommand{\com }[2]{\qty[ #1 ,\, #2 ]}
\newcommand{\trev}{{\hat\Theta}}
\newcommand{\trans}{\mathfrak T}
\newcommand{\GRof}[1]{\hat G^\text{(R)}\qty( #1 )}
\newcommand{\GAof}[1]{\hat G^\text{(A)}\qty( #1 )}
\newcommand{\GaLof}[1]{\hat \Gamma_\text{L}\qty( #1 )}
\newcommand{\GaRof}[1]{\hat \Gamma_\text{R}\qty( #1 )}
\newcommand{\Tmat}{\mathbb T}
\newcommand{\Bmat}{\mathbb B}
\newcommand{\Imat}{\mathbb I}
\newcommand{\Pmat}{\mathbb P}
\newcommand{\Smat}{\mathbb S}
 \newcommand{\imo}{\mathfrak M}
\newcommand{\cc}{\hat c^\dagger}
\newcommand{\ca}{\hat c^{\phantom\dagger}}
\newcommand{\bc}{\hat b^\dagger}
\newcommand{\ba}{\hat b^{\phantom\dagger}}

\begin{document}

\title{Helical orbitals in electrical uni-directional molecular motors}

\author{Štěpán Marek}
\email{stepan.marek@physik.uni-regensburg.de}
\affiliation{Department of Condensed Matter Physics, Faculty of Mathematics and Physics, Charles University, Ke Karlovu 5, 121 16, Praha 2, Czech Republic}
\affiliation{Institute of Theoretical Physics, University of Regensburg, Universitätstraße 31, 93053 Regensburg, Germany}
\affiliation{Regensburg Center for Ultrafast Nanoscopy, University of Regensburg, Universitätstra\ss{}e 31, 93053 Regensburg, Germany}
\author{Wulf Wulfhekel}
\affiliation{Institute for Quantum Materials and Technologies, Karlsruhe Institute of Technology, 76021 Karlsruhe, Germany}
\author{Ferdinand Evers}
\email{ferdinand.evers@physik.uni-regensburg.de}
\affiliation{Institute of Theoretical Physics, University of Regensburg and Halle-Berlin-Regensburg Cluster of Excellence CCE, Universitätstraße 31, 93053 Regensburg, Germany}
\author{Richard Korytár}
\email{richard.korytar@physik.uni-regensburg.de}
\affiliation{Department of Condensed Matter Physics, Faculty of Mathematics and Physics, Charles University, Ke Karlovu 5, 121 16, Praha 2, Czech Republic}


\begin{abstract}
The generation of unidirectional motion has been a long-standing challenge in engineering of molecular motors.  
Here, a mechanism driving the rotation is presented based on electron current through helical orbitals on a $\pi$-bonded carbon chain. 
Such electron current through helical orbitals has been shown to be circulating around the carbon chain. It is 
natural to expect that the associated electronic angular momentum drives a rotation when the current is turned on.
As intuitive as this relation might seem, it is also incomplete because a formal 
definition of helicality in terms of a physical observable has not yet been given.
Such a definition is proposed here. 
Based on this definition, we show how helicality determines the motor's sense of rotation. 
We exemplify the relation between helicality and angular momentum in H\"uckel
models of linear carbon chains (cumulenes and oligoynes).
We attribute the previously reported opposite helicality sense of frontier orbitals (HOMO and LUMO)
to the approximate sub-lattice symmetry.
For oligoynes, this symmetry is hidden in the sense
that it does not reduce to a mere labeling of atoms.
Sub-lattice symmetry, combined with time-reversal invariance, allows us to derive 
Onsager-type reciprocal relations of various linear response coefficients,
dictating
\eg{} an odd energy dependence of angular momentum
response to voltage bias. We propose an observable
consequence of the approximate sub-lattice symmetry:
If the carbon chain is employed as an axle of a molecular rotor, the sense of rotation
is independent on the direction of the current.
\end{abstract}

\maketitle

\section{Introduction}
Molecular motors are molecules that can perform controlled and 
directed rotational motion,
powered by light, an electrochemical 
environment or a chemical reaction
\cite{StoddartNobel,Chuyang2016,Balzani2000,Kassem2017}. The rotational mechanism usually
results from the following simplified picture:
The rotational angle $\theta$ (a slow degree
of freedom), is constrained to an equilibrium value $\theta_0$ by a potential, $V_0(\theta)$,
hindering the rotation.
The excitation induces a fast change of the
potential to $V_1(\theta)$, followed by a relaxation to a new equilibrium,
$\theta_1$. The directionality emerges because one of these potentials 
is not symmetric with regards to clockwise or anti-clockwise rotation
sense (a ``ratchet'') \cite{Julicher1997,Hanggi2009}.

In recent years the tip of a scanning-tunneling microscope
has been used to
wire molecules into a circuit and drive their rotation by the
\revision{tunneling}{electric} current, $I(t)$
\cite{Simpson2019,Ren2020,JasperToenniesRotation,
Stolz2020,Eisenhut2021,Schied2023,
Ho2023,Srivastava2023,Zhang2016,Auyeung2023,Li2025}.
\revision{}{For example, it was reported that $I(t)$ switches in time
through three fixed levels $I_\mathrm A, I_\mathrm B, I_\mathrm C$ such that
the sequence ABC occurs much more often than ACB 
\cite{Stolz2020, SkolautThesis, ExperimentalPaper}. This switching was attributed
to a directional rotation (as opposed to a stochastic motion) and corroborated by
scanning-tunneling maps \cite{Stolz2020}.}
The current-induced rotation has been usually described in terms of
inelastic electron-vibrational transitions \cite{Lorente2020,Ribetto2022}
or electronic current-induced wind-forces \cite{Lu2012,Bustos2013,Hopjan2018,Ribetto2022}. Until now,
 such descriptions have not invoked a specific orbital topology.

Here we explore a
driving force that utilizes directly the angular momentum and helicity of
electrons transported along the motor's axle.
Recently, this concept has been 
explored on the classical level by Koryt\'ar and Evers \cite{KorytarMomentum}.
The authors used a model with
a particle moving along a narrow helical tube.
The current of such particles gave rise to a directed rotation of the tube
due to angular momentum conservation.
In this work we extend this idea
to the quantum-mechanical world. We show
that helical orbitals on a $\pi$-bonded carbon chain can
act analogously to a helical trajectory.

\begin{figure}[h!]
\includegraphics[width=\columnwidth]{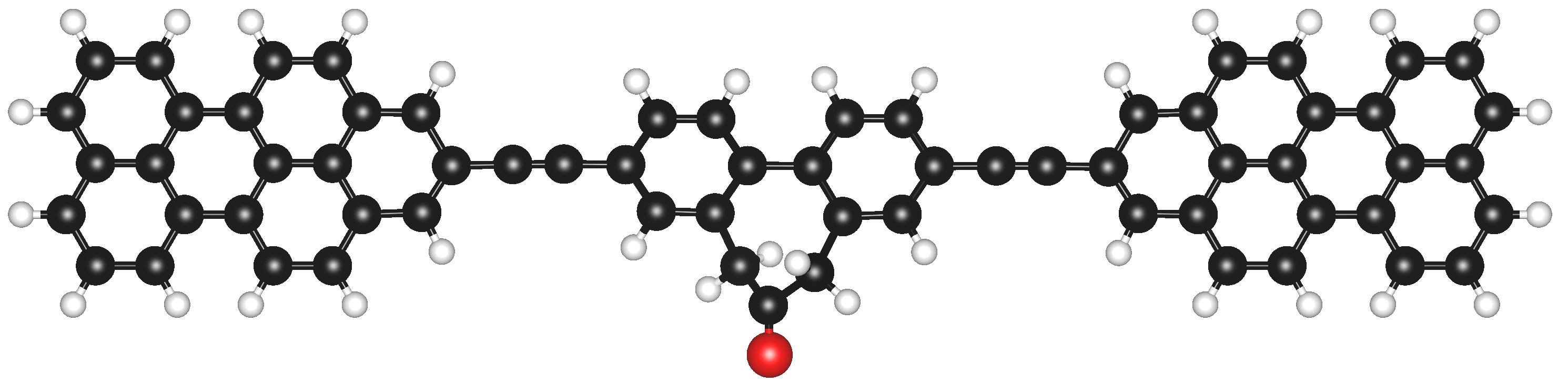}
\includegraphics[width=\columnwidth]{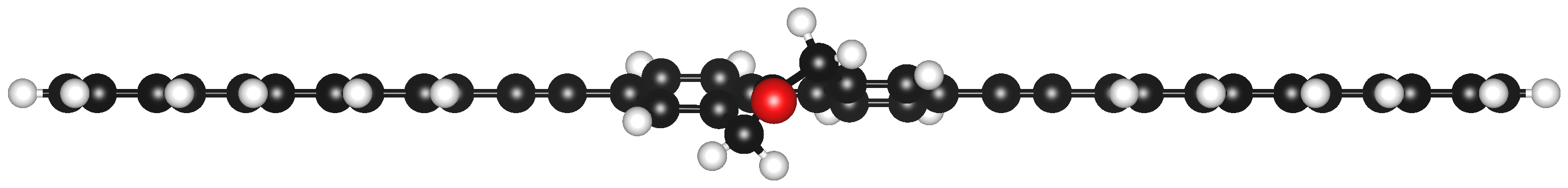}
\includegraphics[width=0.45\columnwidth]{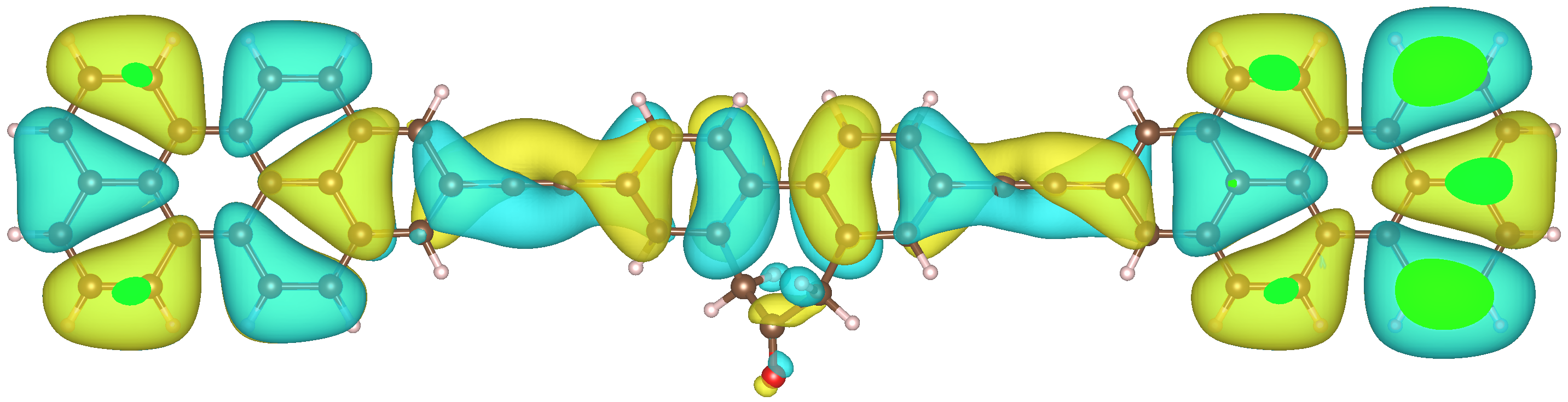}
\includegraphics[width=0.45\columnwidth]{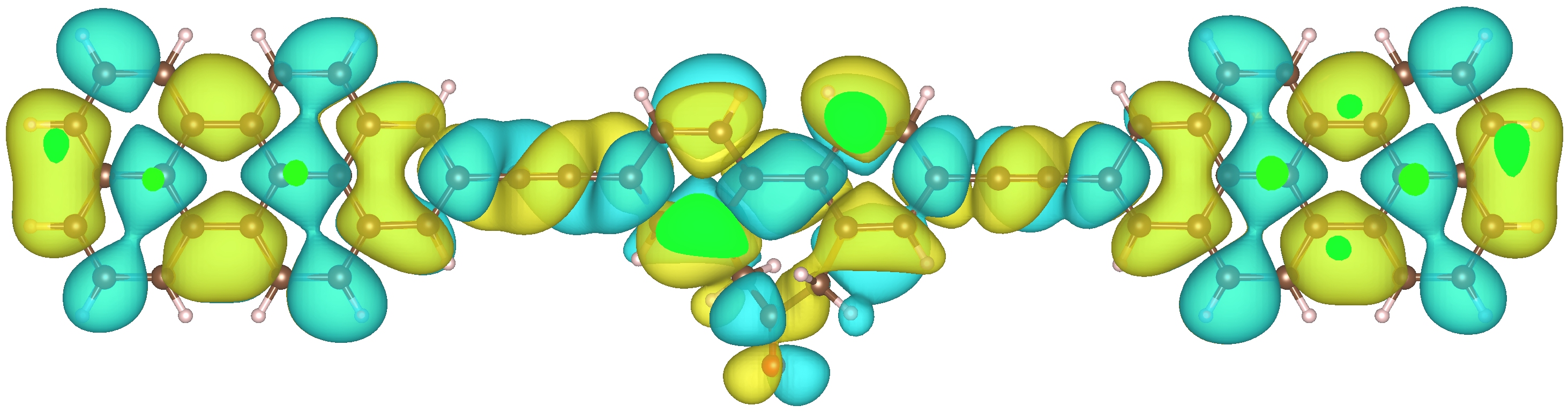}
\includegraphics[width=0.45\columnwidth]{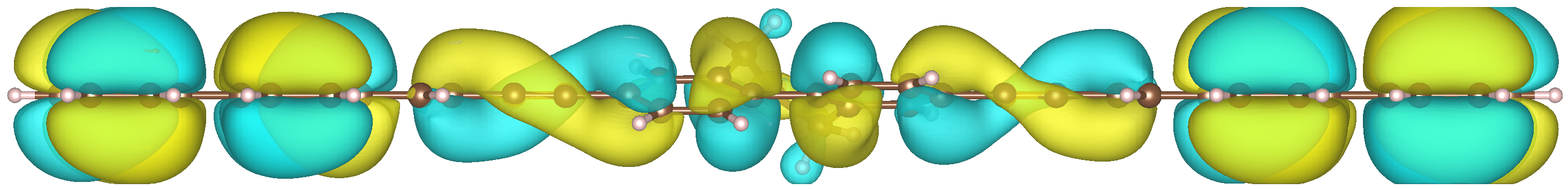}
\includegraphics[width=0.45\columnwidth]{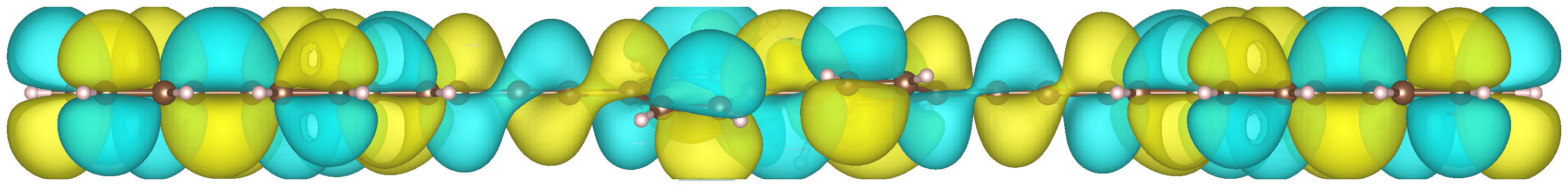}
\caption{\label{f1}Top: Atomic structure of a molecular motor,
containing a central rotating moiety and left and right stators 
(C=gray, H=white, O=red).
The pair of connecting --C$\equiv$C-- chains are molecular axles and bearings. The rotor's
hexagonal rings are tilted by $\approx 29^\circ$,
because of the oxygen
bridge. Each axle is attached to two hexagons that can be tilted with respect to each
other, giving rise to helical orbitals.
Bottom: isosurface of the highest occupied molecular orbital (left) and the lowest
\revision{un}{}occupied molecular orbital (right) calculated in DFT, showing helical
orbitals along the axles with opposite winding senses. The bottom row
portrays these orbitals from a side view.}
\label{abInitio}
\end{figure}

Helical orbitals naturally reside in many $\pi$-bonded carbon chains 
\cite{Hendon2013}. The simplest examples are
cumulenes with end-groups that lower the chain symmetry down to C$_2$
or C$_1$ (axial chirality).
When these chains bridge a molecular junction with a voltage
bias, the electric current winds around
the chain in a helical fashion, as shown by Garner \ea{} \cite{Garner2019}.
Inspired by Skolaut \cite{SkolautThesis}, we employ the linear carbon chain
as an axle connecting rotor and stator moieties of a molecular motor.
\figref{f1} portrays a sample structure and its frontier orbitals.
We argue that the steady-state current flowing through such a molecular architecture is accompanied
by a finite expectation value of electronic angular momentum with respect
to the axle. We demonstrate this by evaluating both helicality and
angular momentum of carbon chains in non-equilibrium.

To this end, we employ a Hückel model 
\cite{KohanoffComputational} of the
$\pi$-orbitals of the carbon chain. The model is applicable
both to cumulenes (carbon chains
with consecutive double bonds) and oligoynes (alternating
single and triple bond). H\"uckel modeling
was proven to exhibit all essential
features of helical orbitals
as shown by Garner \ea{} by comparing with \ai{} calculations \cite{Garner2018}.
Helicality of an orbital can be quite generally quantified
as an average angle of turning of the $\pi$-orbital over one C--C bond.
In this spirit, helicality has been evaluated for particular H\"uckel or
\ai{} wavefunctions \cite{Garner2018,JorgensenSolomonHelical}.
We go beyond these works by introducing a Hermitian operator $\hat h$.
The sign of the expectation value $\bra{\psi}\hat h\ket{\psi}$ determines the clock-wise or anti-clockwise
winding sense of an orbital $\ket\psi$. For weakly-varying wavefunctions,
the
$\hat h$ delivers an average sine of the twisting angle directly.

Promoting helicality to an operator $\hat h$ is advantageous for several reasons:
(1) formally, helicality becomes an observable.
(2) helicality can be evaluated not only for isolated chains but also
   for embedded chains attached to electrodes (molecular junctions),
in non-equilibrium.
(3) in the continuum limit, helicality adopts a transparent form,
$\hat h \propto \hat l\hat p$,
where $\hat l$ is orbital angular momentum of transverse motion and 
$\hat p$ linear momentum along the chain.
The operator product $\hat l\hat p$ is a \textit{helicity} of
a quasi one-dimensional motion. Helicity, in kinematics, is a quantity defined as
the projection of the angular momentum to the direction of motion.
Replacing the orbital $\hat l$ by the spin component, helicity plays
an essential role as an invariant of free Dirac fermions.
Electronic helicity (termed chirality if excitations are gapless) appears as a conserved
quantity in edge states of quantum spin-Hall devices \cite{Koenig2008},
in Weyl semimetals \cite{Armitage2018}
and it has been proposed to generate spin currents in helical molecular junctions
\cite{Liu2021,Korytar2024}.
Our result therefore clarifies a connection between helicality, understood as 
a winding of molecular orbitals and helicity (chirality).
Functional applications of electronic chirality in quantum materials are being
actively researched \cite{Hasan2021,Yan2024}.

We demonstrate that both types of carbon chains -- cumulenes and oligoynes -- are
endowed with a sub-lattice (SL) symmetry. The latter plays an important role
in bi-partite lattices. For example, planar sp$^2$-hybridized 
carbon graphenoid structures
(ribbons, flakes, etc.)
encompass two triangular sub-lattices \cite{CastroNeto2009}. When the electronic structure of
such systems is modeled by nearest-neighbor tight-binding Hamiltonians, the
salient eigen-energies arrange symmetrically around the band center 
(see reviews \cite{Katsnelson2020,Evers2020,Tsuji2018}).
In theoretical chemistry this is known
as the Coulson-Rushbrooke theorem \cite{Coulson1940,Mallion1990}.
The operator $\hat P$
that maps an eigenstate to its mirror partner (in energy space)
merely assigns a minus sign to the wavefunction values on one of the sub-lattices.
Unlike standard symmetries \cite{Altland1997,Koutecky1966}, which are expressed by commutation with the Hamiltonian $\hat H$,
the SL symmetry is formulated by an anti-commutator: $\hat H\hat P + \hat P\hat H=0$.
Consequently, there is no conserved quantity, but a symmetry of eigen-energies and respective eigen-vectors.
Perhaps the earliest occurrence of this 'non-standard' symmetry in quantum physics
was the mass-less limit of the Dirac equation \cite{Schwabl2008}. In such contexts,
the symmetry is called \textit{chiral} \footnote{In the Dirac equation, the 
term chiral refers to spin-momentum locking.}, but this label should not be confused with
structural chirality.

The presence of SL symmetry in sp-hybridized carbon chains is
less obvious for two reasons:
First, unlike in planar sp$^2$ hybridized carbon matter, 
there are two p orbitals per carbon in the $\pi$ system of the chain. Second, the chains
are terminated by end-groups (\eg{} methyl, H, H$_2$) that have a decisive role
in the formation of helical eigenstates. We show that cumulenes have a bi-partite lattice
that
has a rather familiar form: labeling carbons based on their parity along the chain. In the case
of oligoynes the SL symmetry is not apparent from the Hamiltonian matrix; we say 
it is hidden. 
As we show, SL symmetry enforces opposite helicality between 
each molecular orbital and its SL mirror partner.

We employ non-equilibrium Green's function method to evaluate observables
of the chain coupled to leads with bias voltage. The SL symmetry
imposes certain conditions on linear response functions
of helicality and angular momentum. These newly derived relations belong to
Onsager reciprocities since they involve an operation of time-reversal (micro-reversibility).
Most importantly, the response coefficient of angular momentum (with respect to small bias)
is anti-symmetric with respect to the chain's band center.

We find that the expectation value of
angular momentum essentially mirrors the behavior of the orbital winding when
the level broadening is larger than the molecular level spacing.
Finally, we discuss the implications of the angular momentum hosted by helical
orbitals for molecular motors and rotors.

%
%
\section{\label{sec:model}Model Hamiltonian and observables}
The model in which we study the relation of helicality and angular momentum
is described in this section. We introduce the Hamiltonian and define
the respective observables: the angular momentum and helicality operators.
We also describe the coupling to the leads
and finally show that the helicality operator attains a transparent form in the
continuum limit.

\subsection{Model}
We employ a H\"uckel (tight-binding) model to represent 
linear chains of sp-hybridized carbon atoms.
We start with the simpler case of C chains with hydrogens as end-groups.
To induce helical eigenstates, we introduce some more complicated end-groups in
Sec.~\ref{sec:end-groups}, describing methyl-terminated oligoynes and cumulenes.

\subsubsection{Hamiltonian of H-- terminated oligoynes}
\begin{figure}
    \centering
    \includegraphics[width=\columnwidth]{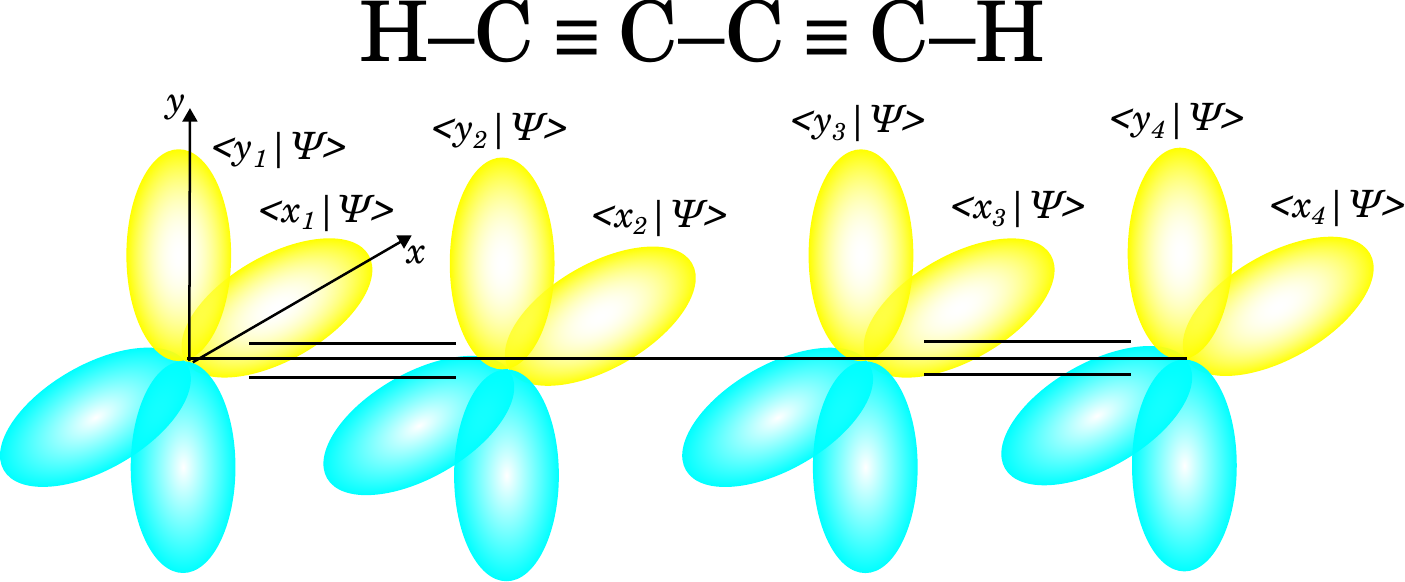}
    \caption{Top: Diacetylene (butadiyne), a member of the oligoyne family with four carbons.
    Bottom: The H\"uckel wave-function of the $\pi$-orbitals is represented in the basis
    of p$_{x,y}$ orbitals on each carbon.\label{modelIllustration}}
\end{figure}

First we start with the Hamiltonian of oligoynes: sp-hybridized carbon chains with H-- end-groups
(see \figref{modelIllustration}).
The basis set of the tight-binding model contains
a p$_x$ and a p$_y$ state on each carbon; $z$ is the chain axis.
The basis is labeled and ordered as follows
\begin{align}
\label{eq:basis}
    \begin{pmatrix}
        \ket{x_1} & \ket{y_1} & \ket{x_2} & \ket{y_2} & \ldots \\
    \end{pmatrix} \: .
\end{align}
where $\ket{x/y_n}$ stands for the p$_{x/y}$-orbital on the $n^\text{th}$ carbon, respectively.
The Hamiltonian of a chain with $N$ carbons has a transparent bra-ket form
\begin{align}
    \hat{H}_0 \coloneqq &{-}t \sum _ {n=1,3,\ldots}^N \bigl[ \ket{x_n}\!\bra{x_{n+1}} + \ket{y_n}\!\bra{y_{n+1}} +
    \text{h.c.} \bigr]\\
     &{-}t' \sum _ {n=2,4,\ldots} ^ {N} \bigl[ \ket{x_n}\!\bra{x_{n+1}} + \ket{y_n}\!\bra{y_{n+1}} +
    \text{h.c.} \bigr] \: ,
\label{eq:h0braket}
\end{align} 
where $-t$ and $-t'$ are the tunneling matrix elements (resonance integrals or hopping elements)
and $\text{h.c}$ stands for Hermitian conjugate.
The difference in $t$ and $t'$ is due to dimerization: In the above equation,
$-t$ pertains to bonds 1--2, 3--4,$\ldots$ and $-t'$ to 2--3,4--5, etc.
The on-site energy has been put to zero.
 Because of
symmetry, the p$_x$ states do not couple to the p$_y$ states.
As a consequence, $\hat{H}_0$ is a direct sum of two identical nearest-neighbor coupled chains.
The eigenstates are not helical, because the system owns a C$_\infty$ axis of rotation.

It will be useful to represent some operators as direct products.
The basis can be re-labeled as $\qty( \ket{\alpha, n})$
with the site index $n$ and $\alpha = \pm 1$ for p$_{x/y}$.
In the case $t=t'$, which will be discussed from section \ref{sec:obs} onwards,
the matrix elements $\bra{\alpha,n}H_0\ket{\alpha',n'}$ can be represented
as
\begin{align}
    -t \left ( \Tmat +\Tmat ^ \top\right ) \otimes \s{0} \: .
    \label{eq:hamUnperturbed}
\end{align}
The matrix $\Tmat$ adopts the indices $n,n'$ and $\s{0}$, the $2\times 2$ 
identity matrix, the indices $\alpha,\alpha'$.
The $N\times N$ matrix $\Tmat$ is given by
\begin{align}
\label{eq:tn}
\Tmat \coloneqq 
\left(
    \begin{array}{ccccc}
        0 & 1  & 0      & \ldots & 0 \\
        0 &  0 & 1      & \ddots & \vdots \\ 
          &    & \ddots & \ddots & 0  \\
        & & & \ddots & 1 \\
        &&&&0
    \end{array}
\right).
\end{align}

\subsubsection{\label{sec:end-groups} Effect of end-groups in cumulenes and oligoynes}
\begin{figure}
\centering
\includegraphics[width=0.8\columnwidth]{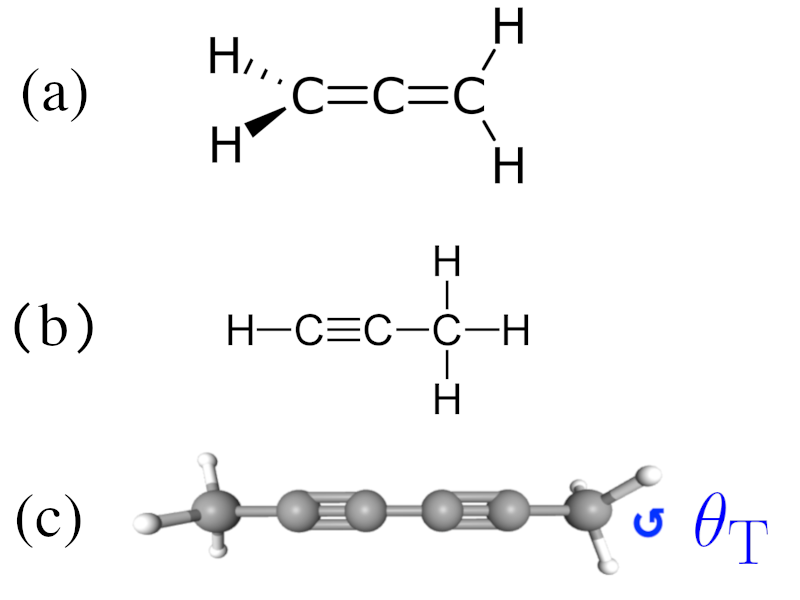}\\
\caption{\label{fig:chainexamples}Examples of carbon chains with different end-groups.
(a) Propadiene with sp$^2$ hybridized end carbons, belongs to allenes. Longer variants, with additional double-bonded carbons inserted 
    in the chain, are called cumulenes.
(b) Propyne can be seen as an oligoyne with a substituted methyl group.
(c) 2,4-hexadiyne with two methyl end-groups, belongs to oligoynes. 
We treat the relative angle
$\thetat$ of
the two groups as a parameter. In cumulenes and 
allenes, carbons are double-bonded. Single and triple bonds alternate
in oligoynes.
}
\end{figure}

To lower the chain symmetry, we consider here replacing the H-- groups by \eg{} a methyl H$_3$C--
or a hydrogen pair, H$_2$--, see \figref{fig:chainexamples}. We assume that the
group consists of sigma bonds only, which are decoupled
from the $\pi$ system of the C chain and well energetically separated from it,
so that these states can be truncated.
The Hamiltonian of a carbon chain with such groups can be written as
\begin{subequations}
\label{eq:fullh}
\begin{align}
\hat H &\coloneqq \hat H_0 + \hat H_{1} + \hat H_{N}\\
\hat H_{1} &\coloneqq \sum_{\alpha,\alpha'} h^{(1)}_{\alpha,\alpha'} \ket{1,\alpha}\!\bra{1,\alpha'}\\
\hat H_{N} &\coloneqq \sum_{\alpha,\alpha'} h^{(N)}_{\alpha,\alpha'} \ket{N,\alpha}\!\bra{N,\alpha'} \: ,
\end{align}
\end{subequations}
with $\hat H_0$ from \epref{eq:h0braket} and the
Hermitian matrices $\hat h^{(1,N)}$ representing the end-groups
and the consecutive loss of the C$_\infty$
symmetry on the sites 1 and $N$.

Having in mind applications of carbon chains as bridges in molecular junctions,
and, most importantly, as axles in molecular motors, we consider
two end-groups rotated with respect to each
other by a dihedral angle $\thetat$.
This motivates us to relate the end-group terms by
\begin{subequations}
\label{eq:hend2}
\begin{align}
\hat h^{(N)} &= \hat R(\thetat)
\hat h^{(1)}
\hat R(-\thetat) \: ,
\end{align}
where the rotation matrix is
\begin{equation}
\hat R(\thetat) = \begin{pmatrix}
\cos(\thetat) & -\sin(\thetat)\\
\sin(\thetat) & \phantom{-}\cos(\thetat)
\end{pmatrix}
\end{equation}
\end{subequations}
and matrix product was employed.

The Hamiltonian \pref{eq:fullh} captures distinct types of end-groups. In this
work we focus on two types of end-groups:
\begin{description}
\item[Model 1 -- cumulenes]For example, propadiene, \figref{fig:chainexamples}(a). 
Here the end carbons are sp$^2$ hybridized ($N=3$).
Thus, one of the p$_{x,y}$ orbitals at the end carbons drops out
from the Hilbert's space \pref{eq:basis}. This can be achieved
by the matrix
\begin{equation}
\label{eq:hend}
\hat h^{(1)} = \begin{pmatrix}\epsilon & 0
\\ 0 & 0 \end{pmatrix},\ \text{and taking }  \epsilon\to\infty.
\end{equation}
Here, $\thetat = \frac\pi 2$.
For $N>3$ these compounds belong to cumulenes.
The Hamiltonian with the 
matrices (\ref{eq:hend},\ref{eq:hend2}) is equivalent to the Hamiltonians investigated 
thoroughly in Refs.~\cite{GunasekaranHelical,Garner2018}.
We revisit this model because our analytical framework allows to rationalize and explain
some numerical results presented in those earlier works.

\item[Model 2 -- oligoynes] This model captures chains with each
end-group of the H$_3$C-- type, as \eg{} in propyne in \figref{fig:chainexamples}(b)
or in 2,4-hexadiyne in \figref{fig:chainexamples}(c).
The carbons of the H$_3$C-- groups are sp$^3$ hybridized. Their effect on
the neighboring carbon sites is accounted for by the Hermitian matrices $\hat h^{(1,N)}$,
small in magnitude in comparison to $t$. 
For general $\thetat$ helical eigenstates
can be expected due to axial chirality \cite{Hendon2013}. To investigate qualitatively the occurrence of helicality with
such boundary terms, we simplify the matrix $\hat h^{(1)}$ to the Ansatz
\begin{equation}
\label{eq:h1simple}
\hat h^{(1)} = \begin{pmatrix} 0  & 0
\\ 0 & \epsilon \end{pmatrix},
\end{equation}
with $0< |\epsilon| \ll t$.
With this parameterization, the matrix commutator
$\qty[\hat h^{(1)}, \hat h^{(N)}]$ is non-vanishing,
except for $\thetat = 0, \pm \frac\pi 2,\pm \pi,\ldots$,
which implies a certain point symmetry. 
\end{description}

Most of this work employs \textbf{Model 2}, and 
\textbf{Model 1} (distinguishing itself by a large $\epsilon$) is invoked only when
stated explicitly.
In both cases, it will be advantageous to express the end-group terms
in a general form
\begin{subequations}
\label{eq:hpauli}
\begin{align}
\hat h^{(1)} &= a^{(1)} \s{0} + b^{(1)} \s{1} + c^{(1)} \s{3}\\
\hat h^{(N)} &= a^{(N)} \s{0} + b^{(N)} \s{1} + c^{(N)} \s{3}.
\end{align}
\end{subequations}
We note that the $\s{2}$ term is absent due to time-reversal
invariance and the $\s{0}$ terms do not select any spatial direction.
Using the $N\times N$ matrices (projectors on the end carbons)
\begin{align}
\label{eq:projectors}
\Bmat_1 \coloneqq 
\left(
    \begin{array}{ccccc}
        1 & 0 & \cdots\\
        0 & 0 & \cdots\\ 
  \vdots  & \vdots & \ddots
    \end{array}
\right),\quad 
\Bmat_N \coloneqq 
\left(
    \begin{array}{ccccc}
  \ddots  & \vdots & \vdots \\
  \cdots  & 0      & 0 \\
  \cdots  &    0   & 1 
  \end{array}
\right)
\end{align}
the end-group terms in \pref{eq:fullh} can written in a product form
\begin{subequations}
\label{eq:hamproductform}
\begin{align}
\hat H_1 &= \Bmat_1 \otimes \hat h^{(1)}\\
\hat H_N &= \Bmat_N \otimes \hat h^{(N)}.
\end{align}
\end{subequations}


\subsubsection{Coupling to the leads}

To carry our calculation over to the non-equilibrium
caused by coupling to the leads and the voltage drop, we 
introduce the
self-energy $\hat\Sigma(E) = \hat\Sigma_\text L(E) + \hat\Sigma_\text R(E)$.
Our analytical results are pursued with only minimal symmetry-imposed constraints
on the form of $\hat\Sigma(E)$, to be specified later.

When chains are bridged across a molecular junction, the left and right lead
couple to the chain's respective end, \ie{}
site $1$ and site $N$.
For our numerical analysis, we adopt a simplified coupling model,
represented by self-energy matrices
\begin{subequations}
\label{eq:sigmas}
\begin{align}
\hat \Sigma_{\text L} &= -\iu\eta \Bmat_1 \otimes \s{0},\\
\hat \Sigma_{\text R} &= -\iu\eta \Bmat_N \otimes \s{0},
\end{align}
\end{subequations}
where \epref{eq:projectors} has been used.
This model assumes that both p orbitals at the end carbons couple equally to the
leads (cylindrical symmetry) with absorption rate $\eta$ ('wide-band').

The 
left and right boundary is coupled to reservoirs with zero-temperature
Fermi distributions 
$f_\text{L/R}(E) = \Theta(\mu_\text{L/R}-E)$ and respective
chemical potentials $\mu_\text{L/R}$. The difference 
$\mu_\text{L}-\mu_\text{R} = eV$ is the external voltage bias.

\subsection{\label{sec:obs}Observables}
In this section we formulate expressions for the angular momentum,
electric current
and helicality operators within the H\"uckel model described previously.
We explain the calculation of their expectation values in non-equilibium,
present some commutaion relations and perform the continuum limit
of the helicality operator.

\subsubsection{Recap: Green's function formalism}
\paragraph*{Green's functions.}
Before we formulate the relevant operators, we digress to
recall the technique for the evaluation of observables in
presence of two reservoirs (following \cite{aitranss,Donarini2024}).
It is advantageous to express
expectation values of
operators using non-equilibrium Green's functions \cite{Kadanoff2018}.
For a one-body operator $\hat O$ (matrix elements
$\bra{n\alpha}\hat O\ket{n'\alpha'}$),
its expectation is given by
\begin{equation}
O(\EF, V ) \coloneqq \expval{\hat O}_{\EF,V} = 
-\iu \int\frac{\dd E}{2\pi} 
\text{Tr} \qty[\hat O \hat G^{<}(E)]
\label{eq:negf}
\end{equation}
where $\hat G^{<}(E)$ denotes the lesser 
Green's function
\begin{multline}
\label{eq:gless}
\hat G^<(E) = \iu \GRof{E}
\Bigl[ \GaLof{E} f_\mathrm L(E) + \\
+ \GaRof{E} f_\mathrm R(E) \Bigr]
\GAof{E}.
\end{multline}
Matrix multiplication is implied.
The advanced (A) and retarded (R) Green's functions 
are defined by
\begin{align}
\GRof{E} &\coloneqq \qty[
E \hat 1 - \hat H - \hat\Sigma(E)]^{-1} \label{eq:GR}\\
\GAof{E} &\coloneqq \qty[\GRof{E}]^\dagger.\label{eq:GA}
\end{align}
We note that in the above expressions, the quantities with hats
should be understood as matrices; likewise $\hat 1$ is a unit matrix.
Finally, the definitions
\begin{align*}
	\hat{\Gamma}_\text{L,R}(E) &= -\iu
\qty(\hat \Sigma_\text{L,R}^\dagger(E) {-} \hat\Sigma_\text{L,R}(E))\\
	\hat{\Gamma}(E) &= -\iu
\qty(\hat \Sigma^\dagger(E) {-} \hat\Sigma(E))
\end{align*}
complete the recapitulation of Green's function formulae.

\paragraph*{Equilibrium and linear response.}
\epref{eq:gless} can be reshaped to 
contain expressions symmetric and anti-symmetric in the Fermi functions,
\begin{subequations}\label{eq:gless2}
\begin{align}
\hat G^<(E) &= \frac\iu 2 \GRof{E}
\qty[ \GaLof{E} + \GaRof{E}] \GAof{E}\cdot \\
&{ }\qquad\qquad \cdot\qty[f_\mathrm L(E) + f_\mathrm R(E)]\ + \\
 &+ \frac\iu 2 \GRof{E}
\qty[ \GaLof{E} - \GaRof{E}]\GAof{E}\cdot\\
&{ }\qquad\qquad \cdot\qty[f_\mathrm L(E) - f_\mathrm R(E)].
\end{align}
\end{subequations}
In equilibrium, the odd term above vanishes, since
$f_\text{L,R}(E) = \Theta(\EF -E)$.
Hence,
\begin{align}
O(\EF,0) &= \int_{-\infty}^\EF \text{Tr}\qty[\hat O\hat A(E)] \dd E \: ,
\label{eq:o0}
\end{align}
where we used \epref{eq:negf} and defined the operator
\begin{subequations}
\label{eq:spectralf}
\begin{align}
\hat A(E) &\coloneqq \frac\iu{2\pi}\left [\GRof{E} -\GAof{E}\right] \\
&= \frac 1{2\pi} \GRof{E} \qty[\GaLof{E} + \GaRof{E}]\GAof{E},
\label{eq:spectralfb}
\end{align}
\end{subequations}
whose the diagonal elements are the familiar projected densities of states.

Certain observables vanish in thermodynamic equilibrium on symmetry grounds. We show in 
Sec.~\ref{sec:trevo} that this is the case of angular momentum, for example.
Vanishing of $O(\EF,0)$ requires $\text{Tr} \qty[\hat O \hat A(E)] = 0$ for
all energies. For such observables, the trace in \pref{eq:negf}
is determined only by the odd term in \epref{eq:gless2}. Explicitly,
\begin{multline}
\label{eq:ononlin}
O(\EF,V) = \half \int \text{Tr}
\Bigl\{ \hat O  \GRof{E} \bigl[ \GaLof{E} -\\
-\GaRof{E} \bigr] \GAof{E} \Bigr\}\qty[f_\mathrm L(E) - f_\mathrm R(E)] \frac{\dd E}{2\pi}.
\end{multline}
Expanding the Fermi functions in voltage
\begin{multline*}
O(\EF,V) =  \frac{eV}2 \int \text{Tr}
\Bigl\{ \hat O  \GRof{E} \bigl[ \GaLof{E} -\\
-\GaRof{E} \bigr] \GAof{E} \Bigr\}\delta(E - \EF)\frac{\dd E}{2\pi}
\ + \mathcal O(V^2)
\end{multline*}
yields the linear-response coefficient
\begin{multline}
\evaluated{\pdv{O(\EF,V)}{V}}_{V=0} = \\ = 
\frac{e}{4\pi} \text{Tr}\qty{ \hat O \GRof{\EF} \qty[\GaLof{E} - \GaRof{E}] \GAof{\EF}}.
\label{eq:dodv}
\end{multline}

\subsubsection{Electric current}
The electric current follows from a standard expression\cite{Donarini2024}
\begin{equation}
\label{eq:iv}
I(V) = \frac{2e}h \int \trans (E) \qty[f_\text L(E) - f_\text R(E)]\dd E \: ,
\end{equation}
with the transmission function
\begin{equation}
\label{eq:te}
\trans (E) \coloneqq \text{Tr} \qty[
\GRof{E}\hat{\Gamma}_\text{L}(E)
\GAof{E}\hat{\Gamma}_\text{R}(E)
].
\end{equation}
Expanding the \epref{eq:iv} to linear order in $V$ yields a well-known Landauer
formula for
the differential conductance at $V=0$,
\[
\eval{\dv{I}{V}}_{V=0} =  \frac{2e^2}h \trans(\EF).
\]

\subsubsection{Angular momentum}
In the $\text p_{x,y}$ basis, the $z$-projection of the
angular momentum satisfies the relations
\begin{align}
    \hat L_z \ket{x_n} &= \phantom - \iu \hbar \ket{y_n} \nonumber \\
    \hat L_z \ket{y_n} &= -\iu \hbar \ket{x_n}.
\end{align}
It can be seen, that
\[ 
\bra{n,\alpha}\hat L_z \ket{n',\alpha'} = 
\hbar\,\delta_{nn'} \qty(\s{2})_{\alpha\alpha'}
\]
with the Pauli matrix $\s{2}$.
In the tensor-product notation of \epref{eq:hamUnperturbed},
the matrix elements of $\hat L_z$
read
\begin{align}
   \hat L_z = \hbar \Imat \otimes \s{2} \label{eq:angularMomentum} \: ,
\end{align}
where $\Imat$ is the $N \times N$ identity matrix.

The expectation value of $\hat L_z$ in presence of the reservoirs can
be expanded in voltage
\begin{equation*}
L_z(\EF,V) \coloneqq
\expval{\hat L_z}_{\EF,V} = L_z(\EF,0) + \pdv{L_z}{V}\; (\EF,0) V + \mathcal O(V^2),
\end{equation*}
where the equilibrium term vanishes due to time-reversal invariance [to be proven in
Sec.~\ref{sec:trevo}]. Hence, 
Eq.~(\ref{eq:dodv}) applies and delivers the linear-response coefficient of angular momentum,
\begin{multline}
\evaluated{\pdv{L_z(\EF,V)}{V}}_{V=0} = \\ =
\frac{e}{4 \pi}
\text{Tr}\Bigl\{\hat L_z \GRof{\EF}
\qty[\GaLof{\EF} - \GaRof{\EF}]\GAof{\EF}\Bigr\}.
\label{eq:dlz0}
\end{multline}

\subsubsection{Helicality operator \label{sec:hop}}
The most natural way to quantify helicality of an orbital
is to measure its average winding angle per C--C bond.
This idea was used to quantify helicality of wavefunctions previously
\cite{Garner2018,JorgensenSolomonHelical}.
Here we extend this idea by formulating a 
Hermitian operator,
whose expectation values quantify the winding of the molecular orbital.

\begin{figure}
    \includegraphics[width=0.5\columnwidth]{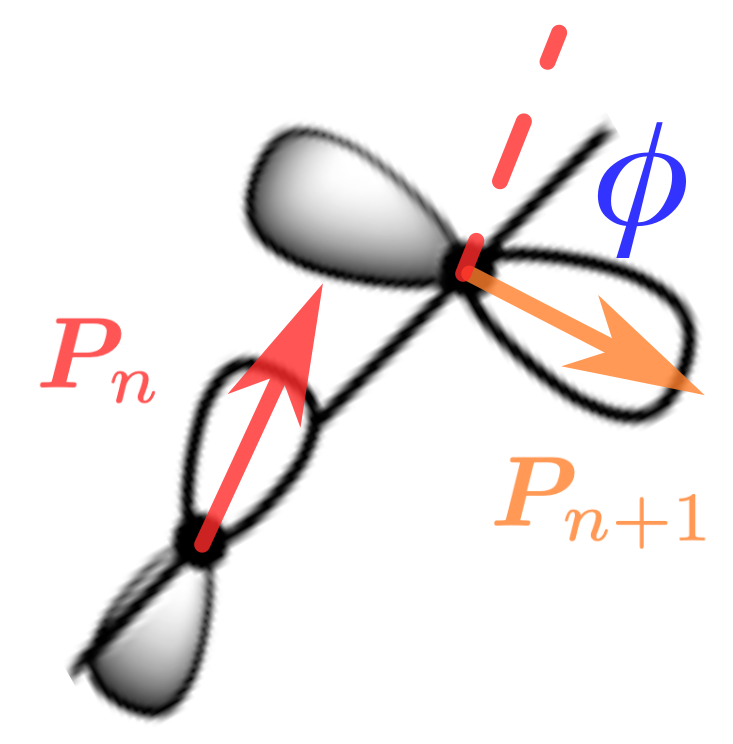}
	\caption{Illustration of the helicality operator. A wavefunction on
	a given site $n$ is a linear combination of p$_x$ and p$_y$
	orbitals. Therefore, it is a p-orbital oriented along a unit vector
	$\bm P_n$. On the consecutive site, the unit vector $\bm P_{n+1}$
	is tilted by an angle $\phi$ with respect to $\bm P_n$. The sign of $\phi$ determines
	the helicality of the given bond, \ie{} the clockwise ($\phi>0$) or
	anti-clockwise winding of the $\pi$-orbital along the bond with increasing $n$. Assuming
	$|\phi| < \pi$, the bond helicality is given by the sign
	of $\bm e_z \cdot \qty(\bm P_n \times\bm P_{n+1})
    = |\bm{P}_n||\bm{P}_{n+1}| \sin\phi$,
    where $\times$ is the cross-product.
	For a given wavefunction, we evaluate $\bm P_n \times\bm P_{n+1}$
	as an expectation value of the local pitch density operator
    $\hat{h}_n$ introduced in the Eq.~\pref{eq:hn}.
}
\label{PolarisationVectors}
\end{figure}

A real wavefunction $\braket{\alpha,n}{\psi}$ (living in the space 
spanned by \pref{eq:basis})
can be represented geometrically
by a series of $N$ 2-dimensional vectors,
\begin{align*}
    \bm P _ n \coloneqq  \begin{pmatrix}
        \braket{x _ n}{\psi} \\
        \braket{y _ n}{\psi} \\
    \end{pmatrix} \:, \quad n=1, \ldots, N 
\end{align*}
pointing perpendicularly to the chain axis ($z$ axis).
The twisting of the orbital is measured by the relative angle of a bond
pair $\qty( \bm P _ n, \bm P_{n+1})$. Vector algebra offers two ways
to extract the angle: a scalar product and a vector product
(or dot and cross products). Garner \ea{} explored the first option
\cite{Garner2018}. Here we take the second one; we have
\[
\bm P_n \times\bm P_{n+1} = \bm e_z \,|\bm{P}_n||\bm{P}_{n+1}| \sin(\phi),
\]
or
\begin{equation}
\label{eq:x}
\tilde h_n \coloneqq 
\bm e_z\cdot \qty(\bm P_n \times\bm P_{n+1}) = |\bm{P}_n||\bm{P}_{n+1}| \sin(\phi),
\end{equation}
where the angle $\phi$ is the relative rotation, see \figref{PolarisationVectors}.
The advantage of \pref{eq:x} is
that $\tilde h_n$ distinguishes between clockwise and anti-clockwise evolution
of vectors. This feature allows us to establish a relation between helicality, defined here,
and angular momentum, in the next section. 

With the definition of the $\times$ product,
\epref{eq:x} turns into
\begin{align}
    \tilde{h}_n = \frac{1}{2} \left ( \braket{x_n}{\psi}^* \braket{y_{n+1}}{\psi} - \braket{y_n}{\psi}^* \braket{x_{n+1}}{\psi} + \text{c.c.} \right ),
    \label{eq:hn_scalar}
\end{align}
where we symmetrized the formula to generalize for complex wavefunctions (c.c. is complex conjugate).
\epref{eq:hn_scalar} can be understood as an expectation value of a bond operator
\begin{align}
    \hat{h}_n = \frac{1}{2} \left ( \ket{x_n} \bra{y_{n+1}} - \ket{y_n} \bra{x_{n+1}} + \text{h.c.}\right ).\label{eq:hn}
\end{align}
We define the helicality operator
by the expression
\begin{align}
\label{eq:hdef}
    \hat{h} = \frac{1}{2} \sum _ {n = 1} ^ {N-1} \left ( \ket{x_n} \bra{y_{n+1}} - \ket{y_n} \bra{x_{n+1}} + \text{h.c.}\right ).
\end{align}
Its expectation values convey the average of the vector products \pref{eq:x} over all bonds.
For $\ket\psi$ that is varying homogeneously along the chain, $|\bm P_n|\approx 1/\sqrt{N}$.
It can be shown, that for long chains
$\bra\psi \hat h\ket\psi$ offers a straightforward
interpretation as an average sine of the consecutive
p-orbitals' turning angle, see \appref{app:averagePitchDensityName}.
In \appref{app:previousWork} we evaluate $\hat h$ for eigenstates
of (\ref{eq:fullh},\ref{eq:hend}) with $\epsilon\to\infty$ obtained
analytically by Gunasekaraman and Venkataraman\cite{GunasekaranHelical}
to model cumulenes.

The product form of the helicality operator reads
\begin{align}
\hat h = \frac{\ci}{2}\left( \Tmat -\Tmat^\top \right) \otimes\s{2} \: ,
\label{eq:averagePitchDensity}
\end{align}
where the matrix \pref{eq:tn} has been used.
Comparing with \eqref{eq:angularMomentum}
we immediately see that the helicality operator commutes with the angular
momentum operator,
\[
\qty[\hat L_z, \hat h] = 0.
\]
The crucial difference between these two observables lies
in the behavior under time-reversal. While $\hat L_z$ changes
sign, the real $\hat h$ is invariant under time-reversal. Explicit
form of the time-reversal operator is shown in Sec.~\ref{sec:basicSym}.

The eigenstates of $\hat h$ are complex, they resemble 
'circularly-polarized'
waves, as shown in \appref{app:eigenvectors}.
In \appref{app:averagePitchDensityCommutator} we show that
the helicality operator commutes with chain Hamiltonian $\hat H_0$ up
to the end-group terms,
\[
\qty[\hat H_0, \hat h] = t(\Bmat_ 1 - \Bmat_ N) \otimes \ci \s{2},
\]
where $\Bmat_N$ and $\Bmat_1$ are defined in \epref{eq:projectors}.
It can be shown that when chains are closed into
rings (periodic boundary condition), the commutator vanishes, and
$\hat H_0$ has common eigenstates with $\hat h$ (\appref{app:pbc}).

It is of interest to evaluate helicality for chains coupled to reservoirs.
Using the notation of \epref{eq:negf},
\begin{equation*}
h(\EF,V)  = h(\EF,0) + \order{V}
\end{equation*}
and formula \pref{eq:o0} yields
\begin{align}
\label{eq:hv0}
h(\EF,0) &= \int_{-\infty}^\EF \text{Tr}\left [ \hat h\, \hat A(E)\right]  \dd E,
\end{align}
and the quasi-equilibrium response coefficient
\begin{equation}
\label{eq:dhde}
\evaluated{\pdv{h(E,0)}{E}}_\EF = \text{Tr}\left [ \hat h\, \hat A(\EF)\right].
\end{equation}
In stark contrast to the angular momentum, the equilibrium value of
helicality can be finite. This should not come as a surprise, given that
in our construction helicality of orbitals in isolated chains can be non-zero.

\subsubsection{Helicality operator in the continuum limit \label{sec:hcont}}
Before presenting extensive numerical analysis, we digress to perform
a continuum limit of the helicality operator in order to lend extra plausibility to
our definition of $\hat h$.
The coordinate along the molecular backbone is chosen to be $z$ and the
carbons are located at $na$, $a$ being the bond length.
The H\"uckel wavefunction of the chain is here denoted by the
column vector $\psi(z)$ with two entries $\psi_x(z)$ and $\psi_y(z)$, which
are the projections on the $\text p_{x,y}$ orbitals on the site $z$.
The right-hand side of \pref{eq:hn_scalar} in this fashion reads
\begin{multline}
    \frac{1}{2} \left [ \psi^*_x(z)\psi_y(z+a) -
      \psi_y^*(z)\psi_x(z+a)
    \,+\, \text{c.c.} \right ] =\\
    = \frac{\iu}{2} \qty[ \psi^\dagger(z)\,\s{2}\, \psi(z+a)\, +\, \text{c.c.}],
    \label{eq:hn_val}
\end{multline}
where matrix multiplication is implied in the second line. We can expand
$\psi(z+a) = \psi(z) + a\partial_z\psi(z) + \order{a^2}$. If the wavefunction
varies smoothly, second and higher-order terms can be neglected;
\epref{eq:hn_val} becomes
\begin{multline}
\frac{\iu a}{2} \qty{\psi^\dagger(z) \s{2} \partial_z\psi(z) - 
\qty[\partial_z\psi^\dagger(z)] \s{2}\psi(z)} =\\
= \frac{\iu a}2 \int 
\bigr[ \psi^\dagger(z') \s{2} \delta(z -z') \partial_{z'}\psi(z') +\\
+ \psi^\dagger(z') \s{2} \partial_{z'} \delta(z-z') \psi(z') \bigl]\dd z'.
\end{multline}
The second line follows from an integration of a Dirac's delta function;
in the third line we used integration by parts.
The integral expression above offers the definition of the helicality density
\begin{align}
\hat h(z) &= \frac{\iu}2 \s{2} \qty{\partial_z, \delta(z-\hat z)}\\
&= -\frac{m}\hbar \s{2} \hat j(z),
\end{align}
where we introduced the longitudinal particle current density operator
$\hat j(z) \coloneqq \half \qty{\frac{\hat p_z}{m}, \delta(z-\hat z)}$,
the anticommutator $\qty{,}$ and the momentum operator
 $\hat p_z \coloneqq -\iu\hbar \partial_z$.

The integral of $\hat h(z)$ leads to a continuum representation of
the helicality operator:
\begin{equation}
\hat h = \int \hat h(z)\dd z = -\frac{1}{\hbar}\, \hat p_z \s{2}.
\end{equation}
Identifying $\hat L_z = \hbar \s{2}$ [cf. \epref{eq:angularMomentum}], we arrive at the
important identity
\begin{equation}
\label{eq:important}
\hat h = -\frac{1}{\hbar^2}\, \hat p_z \hat L_z.
\end{equation}
This expression has the form (up to a factor) of the helicity
operator $\hat p_z \hat L_z$ of a particle constrained to move
along the $z$ direction\cite{RQMSchweber}. The pre-factor contains $\hbar^{-2}$
because $\hat h$, with the continuum limit taken, has dimension 1 over length.
The minus sign should not cause confusion: the left side of \pref{eq:important}
quantifies the wavefunction 
shape and the right side an actual particle motion.

Notice that the $\hat L_z$ can be replaced
by its real-space representation $-\iu\hbar\qty(\hat x\partial_y 
- \hat y \partial_x)$. This allows to employ the helicality operator beyond
the H\"uckel theory, \textit{e.g.} with a richer atom-centered basis. For instance,
the formula \eqref{eq:important} can be utilized in \emph{ab-initio} calculations,
as an alternative to the nodal plane precession technique of Ref.~\cite{JorgensenSolomonHelical}.

Helicity encountered in \epref{eq:important} has a
transparent geometrical meaning for planewaves:
Such a state is helical when it has both
angular and longitudinal momentum nonzero. A classical picture of
a particle moving along a helix can be visualized.
But helicity can be finite in standing waves, too.
Standing waves are composed of pairs of planewaves, related by time-reversal. Formally,
$\psi \propto \qty[\ket{p,\uparrow} + \ket{-p,\downarrow}]/\sqrt 2$,
where the arrows denote the angular momentum projection.
Notice, that
$\expval{\hat L_z}$ and $\expval{\hat p_z}$ vanish for $\psi$.
However, this is not true for $\expval{\hat L_z\hat p_z}$.
A simple calculation shows that helicity is finite in $\psi$, although
the linear and angular momenta vanish.
Algebraically, this behavior is due to  the product 
$\hat p_z\hat L_z$ commuting with the time-reversal operator,
unlike the linear and angular momenta, which anti-commute.
Therefore,
for isolated carbon chains, helical eigenstates (standing waves) have $\expval{\hat h}\ne 0$,
although $\expval{\hat L_z}=0$. Coupling to the reservoirs with steady-state current breaks
time-reversal; consequently $\expval{\hat L_z}_\text{res} \ne 0$.
In the continuum limit, the latter result must hold 
because of \epref{eq:important}.

\section{Symmetry analysis of the Hamiltonian and the observables}
We show in Sec.~\ref{sec:sls} that Models 1 and 2 enjoy a SL 
symmetry and explore its consequences for molecular orbitals of isolated
chains. For chains coupled to leads, we derive anti-symmetry relations of
linear-response coefficients of helicality and angular momentum. These
relations follow from SL symmetry in combination with
time-reversal. The presence of both symmetries -- SL and time-reversal --
implies that both Models 1 and 2 belong to the symmetry class BDI of 
the ten-fold classification \cite{Altland1997,Chiu2016}.


Throughout this section we set $t=t'$ for the sake of simplicity,
although the main result -- the effect of SL symmetry on angular momentum and helicality
 -- remains valid when $t\ne t'$ or when hoppings are inhomogeneous.

\subsection{Basic symmetries of the chains\label{sec:basicSym}}
We start with rather evident symmetries,
the time-reversal invariance (TRI) and rotational symmetry. Implications
of TRI are explored: vanishing equilibrium angular momentum and 
symmetrization of the angular momentum response coefficient.

\subsubsection{Cylindrical symmetry}
Carbon chains terminated by H-- on both ends, such as the one shown in \figref{modelIllustration},
enjoy full rotational invariance, C$_\infty$, around the $z$ axis.
The commutator
\[
\qty[\hat H_0, \hat L_z ] = 0
\]
can be most easily derived from expressions 
(\ref{eq:hamUnperturbed},\ref{eq:angularMomentum}).
The loss of rotational symmetry that comes with end-group terms
of Model 1 and Model 2, 
is a prerequisite for finite expectation value of angular momentum
in chains with current flow.

\subsubsection{Time-reversal}
An important role in the subsequent analysis of transport is
 played by the time-reversal operator $\trev$.
We emphasize that the basis in which the H\"uckel model is written
is real, \ie{} the p-orbitals are real harmonics. In the product
form of operators, such as \pref{eq:angularMomentum}, the Pauli
matrices do not represent spin, but a pseudospin, that does not
invert sign upon time-reversal.
Consequently, the time-reversal operator reduces to
complex conjugation $\mathcal C$,
\begin{equation}
\label{eq:trev}
\trev = \mathbb I\otimes \s{0}\, \mathcal C\ \equiv \mathcal C.
\end{equation}

Hence, the following identities hold
\begin{subequations}
\label{eq:TLh}
\begin{align}
\acom{\hat L_z}{\trev} &= 0, &{ }
\com{ \hat h}{\trev} &= 0,
\end{align}
\end{subequations}
showing a distinctively different behavior of helicality and angular
momentum, already discussed in Secs.~\ref{sec:hop} and \ref{sec:hcont}.

Since the Hamiltonian is real, 
\begin{equation}
\com{ \hat H}{\trev} = 0
\end{equation}
with any end-group type.

To analyze the self-energies, we introduce the decomposition
\begin{subequations}
\label{eq:SigmaDecomp}
\begin{align}
\hat \Sigma(E)  &= \hat \Delta(E) - \frac\iu 2\hat \Gamma(E),\\
\hat \Delta(E) &\coloneqq \half\qty[\Sigma(E) + \Sigma^\dagger(E)],\\
\hat \Gamma(E)  &\coloneqq  \iu
\qty[\hat \Sigma(E) -\hat\Sigma^\dagger(E)],
\end{align}
\end{subequations}
where $\hat\Gamma(E)$ and $\hat \Delta(E)$ are hermitian by definition.
From embedding theory \cite{Donarini2024}, it can be shown that both quantities must be real
if the leads are time-reversal invariant. Therefore,
\begin{equation}
\label{eq:trevGamma}
\com{\hat \Gamma}{\trev} = 0
\end{equation}
and $\trev \hat \Sigma(E) = \hat \Sigma^\dagger(E)\trev$.
The above conditions for $\hat \Sigma(E)$ apply also to the
left and right components, $\hat \Sigma_\text{L,R}(E)$, separately.
Finally \epref{eq:GA} brings us to the condition of time-reversal
of chains coupled to leads,
\begin{equation}
\label{eq:trevg}
\trev \GRof{E} = \GAof{E}\trev.
\end{equation}

\subsubsection{Consequences of time-reversal invariance (TRI)
for observables\label{sec:trevo}}
It is known that 
some observables change sign upon time-reversal, others
don't [\epref{eq:TLh} exemplifies this]. This is reflected, quantum-mechanically, by the (anti-) commutation
with $\trev$,
\begin{equation}
\label{eq:signature}
\trev\hat O = \pm \hat O\trev,
\end{equation}
where the sign is sometimes called a 'signature' \cite{Schwabl2008}.
Let us inspect the equilibrium expectation value \pref{eq:o0}
of an observable $\hat O$.
$\hat A(E)$ is hermitian by definition [see \epref{eq:spectralf}],
but TRI \pref{eq:trevg}
restricts it to be real and symmetric:
\[
\trev \hat A(E) = \hat A(A) \trev.
\]
We insert the identity $\trev^2 = \hat 1$ into the trace \pref{eq:o0}
and apply commutation rules,
\[
\text{Tr}\qty[ \trev\trev\hat O \hat A(E) ] = \pm \text{Tr}\qty[ \trev\hat O \hat A(E) \trev].
\]
The signs correspond to the ones in \epref{eq:signature}. The right $\trev$ can be dropped
as it \revision{has nothing to conjugate}{operates on real wavefunctions} and 
the left one can be replaced by the conjugation of the resulting trace. But the
latter is real, which makes us conclude that
\begin{equation}
\label{eq:trevtr}
\text{Tr}\qty[\hat O \hat A(E) ] = \pm \text{Tr}\qty[ \hat O \hat A(E)]
\end{equation}
and $O(\EF, 0) = \pm O(\EF, 0)$. Consequently, observables that change sign upon time-reversal
have a vanishing expectation value in equilibrium in a TRI system. This is the case for
the angular momentum
\begin{equation*}
L_z(\EF, 0 ) = 0.
\end{equation*}
Helicality, on the other hand, does not vanish when $V=0$.

\subsection{Sub-lattice symmetry in cumulenes \label{sec:slcumu}}

Both models 1 and 2 are endowed with the SL symmetry, which has
important implications not only for helicality, but also for angular momentum.
The SL symmetry has different form in the two cases and we start
with the simpler case, Model 1. Here, the SLs are constructed
according to the parity of each carbon along the chain. Hence, the bipartite lattice in
cumulenes is analogous to the notion of bipartite lattices in planar sp$^2$ hybridized
carbons (``graphenoids''). Both p orbitals on
the same atom belong to the same SL. On the ends, only the p orbital,
pertaining to the $\pi$ system, counts. To demonstrate that algebraically,
we introduce the operator that flips signs on odd-numbered carbons,
\begin{equation}
\label{eq:pdef}
\hat P \coloneqq \sum_{n=1}^N (-1)^n\qty[\ket{y_n}\!\bra{y_{n}} + \ket{x_n}\!\bra{x_{n}}]
 = \Pmat\otimes\s{0},
\end{equation}
where the diagonal matrix
\begin{equation}
\label{eq:P}
\Pmat \coloneqq \left(
    \begin{array}{rrrr}
       -1 &  0 & 0 & \cdots\\
        0 & +1 & 0 & \cdots\\
        0 &  0 &-1 & \cdots\\
  \vdots  & \vdots & \vdots & \ddots
    \end{array}
\right) = \text{diag}\qty(-1,+1,-1,+1,\ldots)
\end{equation}
is defined.
The SL symmetry is a statement that with $\hat H = \hat H_0+\hat H_1+\hat H_N$ and
Eqs.~(\ref{eq:hamUnperturbed},\ref{eq:hamproductform},\ref{eq:hend}) it holds that
\begin{equation}
\label{eq:acomModel1}
\acom{\hat H}{ \hat P} = 0 \quad\text{(Model 1)}
\end{equation}
when $\epsilon \to \infty$.

To prove that,
we first show that the 'bulk' part of the Hamiltonian
anti-commutes with $\hat P$, namely,
\begin{equation}
\label{eq:com0}
\acom{\hat H_0}{\hat P} = 0,
\end{equation}
with \epref{eq:hamUnperturbed}.
This follows from
\begin{equation}
\label{eq:TP}
\acom{\Tmat^{\phantom\top}}{\,\Pmat} = \acom{\Tmat^\top}{\Pmat} = 0
\end{equation}
[see also  \epref{eq:tn}].
Next, the left boundary term,
site 1  with \epref{eq:hend}, delivers the result
\begin{equation}
\label{eq:H1P}
\acom{\hat H_1}{ \hat P} = -2 \epsilon \ket{x_1}\!\bra{x_1}.
\end{equation}
In the limit $\epsilon \to \infty$, wavefunctions of the $\pi$ orbital
system have a vanishing weight on the $\ket{x_1}$ site \footnote{The requirement
of $\lim_{\epsilon\to\infty}\, \epsilon \braket{x_1}{\psi}\, =0$ is made here. It is guaranteed
by the exponential decay of $\psi$ in the tunnel region.}.
Therefore, the relevant
matrix elements of \pref{eq:H1P} vanish. Analogously,
\begin{equation}
\label{eq:H1N}
\acom{\hat H_N}{\hat P} = 2\epsilon(-1)^N \ \Bmat_N\otimes \qty[\hat R(\thetat) \cdot
\begin{pmatrix}
1 & 0 \\ 0 & 0
\end{pmatrix}\cdot
\hat R^\top(\thetat)],
\end{equation}
where Eqs.~(\ref{eq:hend2},\ref{eq:hend},\ref{eq:hamproductform})
have been used. The square braces enclose a $2\times 2$ matrix. It is a projector
on a linear combination of $\ket{x_N}, \ket{y_N}$ states that is truncated from
the $\pi$ system. Again, the \epref{eq:H1N} has vanishing matrix elements within the $\pi$ system, and
\epref{eq:acomModel1} is proven.

It is not hard to show, that
\begin{subequations}
\label{eq:Pcoms}
\begin{equation}
\acom{\hat h}{\hat P} = 0
\end{equation}
using \epref{eq:TP}.
\figref{PolarisationVectors} offers a geometric view of the anti-commutator: The $\hat P$ inverts
the phase of the p-orbitals of one member of the bond. If one of the components of vector $\bm P$ flips sign,
so does the cross-product and $\phi$. Further identities
\begin{align}
 \hat P^\dagger &= \hat P,\\
  \hat P^2 &= 1,\\
 \com{\hat L_z}{\hat P} &= 0 \label{eq:comLP}
\end{align}
\end{subequations}
will be instrumental for deriving consequences of the SL symmetry.
\epref{eq:comLP} says that a simultaneous inversion
of $x$ and $y$ axes does not change the sense of rotation around $z$.

It can be expected that in more complete models of cumulenes, the SL symmetry
holds approximately, only. One example of the symmetry-breaking term is
the on-site potential of the p$_y$ orbital on the first carbon, which
can be
slightly different from the carbons in the middle of the chain. The effect of this
term is small because of the small energy and because of locality. We look at this
in more depth in the case of oligoynes.


\subsection{\label{sec:sls} A $\hat Q$-symmetry in oligoynes}
We show that Model 2 also enjoys an anti-commutator relation of the form 
\pref{eq:acomModel1},  with $\hat P$ replaced by a new operator $\hat Q$.
The meaning of the anti-commutator is less straight-forward than in cumulenes.
In Sec.~\ref{sec:chiral} we show that a suitable basis change
transforms the Hamiltonian into a form where the SL is revealed.

We define an operator
\begin{equation}
\label{eq:qdef}
\hat Q \coloneqq 
    \sum_{n = 1} ^ {N} (-1)^n
     \Bigl [\ket{y_n}\!\bra{x_{n}} - \ket{x_n}\!\bra{y_{n}}  \Bigr ]
\end{equation}
which, up to signs, exchanges locally the p$_x$ and p$_y$ orbitals.
 In the tensor product form,
$\hat Q = -\iu\Pmat \otimes \s{2}$, with \epref{eq:P}.

The central result of this work is that
for the Hamiltonian $\hat H$ with the parametrization \pref{eq:hpauli}
we have
\begin{equation}
\label{eq:com1}
\acom{\hat H}{\hat{Q}} = 2\iu\left(a^{(1)}\Bmat_1 - (-1)^Na^{(N)} \Bmat_N\right)\otimes \s{2} 
\end{equation}
where $\Bmat_{1,N}$ are defined in \epref{eq:projectors}.
Moreover, it holds that
\begin{align}
\label{eq:com2}
\com{\hat L_z}{ \hat Q} &= 0\\
\label{eq:com3}
\acom{\hat h}{\hat Q} &= 0.
\end{align}

Let us start with the 'bulk' term,
\[
\acom{\hat H_0}{\hat Q} = -\iu t\acom{\Tmat + \Tmat^\top}{\Pmat} \otimes \s{2} =0,
\]
using \pref{eq:hamUnperturbed} and \pref{eq:TP}.
Anti-commutation rules $\acom{\s{i}}{\s{j}} = 2\s{0}\delta_{ij}$ $(i>0,j >0)$
imply for the end-group terms
\begin{alignat*}{3}
\acom{\hat H_1}{\hat Q} &= (-1)^1a^{(1)} \Bmat_1 \otimes \acom{\s{0}}{ -\iu\s{2}} &=\\
&= 2\iu a^{(1)} \Bmat_1 \otimes \s{2} \\
\acom{\hat H_N}{\hat Q} &= (-1)^Na^{(N)} \Bmat_N \otimes \acom{\s{0}}{ -\iu\s{2}} &=\\
&= (-1)^{N+1}2\iu a^{(N)} \Bmat_N \otimes \s{2}.
\end{alignat*}
Grouping the three Hamiltonian terms together, we prove \epref{eq:com1}.
Using \epref{eq:TP}, the relation
\begin{align*}
\acom{ (\Tmat - \Tmat^\top)\otimes \s{2}} {\Pmat \otimes \s{2}} = 0
\end{align*}
follows. Therefore,
\epref{eq:com3} holds in view of the definition \pref{eq:averagePitchDensity}.
On the other hand, commutator \pref{eq:com2} holds because
both $\hat L_z$ and $\hat Q$ are proportional to $\s{2}$ and $\com{\Imat}{\Pmat} = 0$
(see \epref{eq:angularMomentum}). The unitarity relations will prove useful
\begin{align}
\label{eq:Qprop}
\hat Q^{-1} &= -\hat Q  =\hat Q^{\dagger},
\end{align}
which follow from idempotency $\s{2}^2 = \s{0}$.

Concluding, the Hamiltonian of Model 2 anti-commutes with $\hat Q$ up to boundary
terms on the right-hand side of \pref{eq:com1}. We refer to the vanishing of these terms 
as SL symmetry of Model 2. Namely,
\begin{equation}
\label{eq:ph}
a^{(1)} = a^{(N)} = 0\quad (\text{\it SL symmetry}).
\end{equation}
The Hamiltonian terms proportional to $a^{(1,N)}$
have the $\s{0}$ form; these terms
offset the on-site potential on the end sites equally for each p orbital. Therefore, they
do not induce helicality and may be seen as irrelevant for
current-induced angular momentum. For the discussion
of the generation of helical orbitals the $\s{0}$ terms can be taken to be zero. 
In Appendix \ref{app:sublattice}
 we show that these terms lead to weak violation
of the SL symmetry, only. 

\subsection{\label{sec:chiral} A sub-lattice representation of Model 2}
\begin{figure}
\includegraphics[width=\columnwidth]{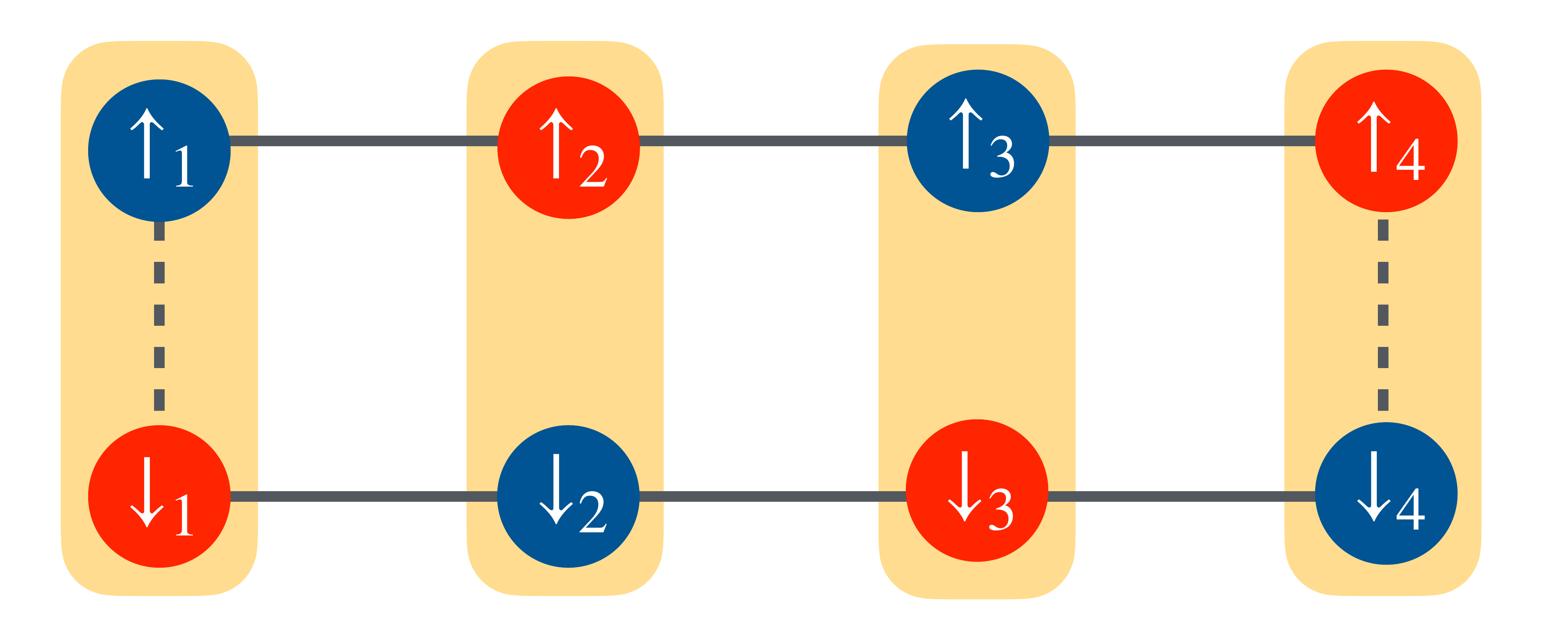}
\caption{\label{fig:ladder} Scheme of the SL symmetric model of
a carbon chain with 4 atoms in the
representation of angular momentum eigenstates. On each atom (yellow box),
 there are
two $\hat L_z$ eigenstates with $L_z = \pm\hbar$, labeled by the arrow. Blue (red) color denotes
the emergent sub-lattice A (B) of the model. Solid lines represent the nearest-neighbor 
'hopping' terms in the Hamiltonian, equal to $-t$. Dashed lines represent \revision{}{the} couplings 
\revision{from}{within}
the end-groups. Atoms of the same SL
do not couple. The couplings effectively arrange into a ring.
}
\end{figure}

We show that a suitable change of basis transforms the the Hamiltonian
to a matrix form which reveals a bipartite lattice structure.
To achieve this, we bring
the operator $\hat Q$ to a diagonal form.
It follows that the transformation diagonalizes $\s{2}$. Incidentally, 
because of \epref{eq:angularMomentum},
this is the transformation to angular momentum eigenstates. Let
\[
\hat U \coloneqq 
\Imat \otimes \exp{-\iu \frac\pi 4\s{1}} = \Imat \otimes
\frac 1{\sqrt 2} \qty(\s{0} - \iu \s{1})
\]
map to a new (complex) basis, that can be labeled 
\begin{equation}
\label{eq:basis2}
\qty(\ket{\uparrow_1}, \ket{\downarrow_1}, \ket{\uparrow_2}, \ket{\downarrow_2},
\ldots )
\end{equation}
(cf. \epref{eq:basis}) where the arrow is a shorthand for the orbital angular momentum 
projection, $\pm \hbar$, and the subscript labels the atom.
Then $\hat U \hat Q \hat U^\dagger = -\iu \Pmat\otimes \s{3}$
is a diagonal matrix. Apart from the factor $+\iu$,
the diagonal elements multiply wavefunction
values on every other point of the new 'lattice' by $-1$.
Specifically, SL B, receiving the minus signs, is seen
to be 
\[
\text{SL B}
= \qty( \ket{\downarrow_1}, \ket{\uparrow_2}, \ket{\downarrow_3}, \ket{\uparrow_4},\ldots )
\]
and SL A is the remainder (see \figref{fig:ladder}).

Next, with \epref{eq:hpauli}, the end-term transforms
\[
\hat U \hat H_1 \hat U^\dagger = \Bmat_1 \otimes 
\qty[a^{(1)}\s{0} + b^{(1)} \s{1} - c^{(1)} \s{2}]
\]
and similarly for $\hat H_N$. Notice that there are no $\s{3}$ terms
after the transformation. 
Moreover, $\hat H_0$ does not
change its matrix form, because it commutes with angular momentum.
Summarizing, the Hamiltonian, including end-terms, transforms to
\begin{multline*}
\hat{\widetilde H} \coloneqq \hat U \hat H \hat U^\dagger = 
-t\qty(\Tmat + \Tmat^\top)\otimes \s{0} + \\
+ \Bmat_1 \otimes
\qty[a^{(1)}\s{0} + b^{(1)} \s{1} - c^{(1)} \s{2}] +\\
+
\Bmat_N \otimes
\qty[a^{(N)}\s{0} + b^{(N)} \s{1} - c^{(N)} \s{2}].
\end{multline*}
 When the condition \pref{eq:ph} holds,
$\hat{\widetilde H}$
is free from matrix elements that operate on basis functions of the same SL
(see \figref{fig:ladder} for a graphical representation).
If the basis set \pref{eq:basis2} is re-ordered so that
all A-lattice elements come before all B-lattice ones, the matrix of $\hat{\widetilde H}$
has the block structure
\begin{equation}
\label{eq:chiralH}
\begin{pmatrix}
0 & \mathbf w \\
\mathbf w^\dagger & 0
\end{pmatrix}
\end{equation}
with only the non-diagonal blocks $\mathbf w, \mathbf w^\dagger$ non-zero.
Concluding, in the representation of $\hat L_z$ eigenstates, the
hidden SL symmetry uncovers itself. 

Hamiltonians of the form \pref{eq:chiralH} are also called
\textit{chiral}, or, possessing chiral symmetry,
in relativistic physics,
in mesoscopic physics of bi-partite lattice systems and other areas of
quantum physics \cite{Evers2008,Chiu2016}.
To avoid confusion, we stress that it is not related to
structural chirality of the molecule.

The helicality operator transforms to
\[
\hat U\hat h \hat U^\dagger = \frac \iu 2 \qty(\Tmat - \Tmat^\top)
\otimes \s{3}
\]
which allows for a transparent view of its eigenstates, namely, as
direct products of angular momentum and 'linear' momentum eigenstates.

\subsection{Sub-lattice symmetric self-energy}
To study the consequences of SL symmetry for non-equilibrium situations with current,
SL symmetry must be \revision{postulated}{assumed} for the self-energy. \revision{To do so, }{}
We begin by exploring the structure of the
Green's function of isolated chains, denoted by $\hat g^\text{(R)}(E)$.
This is \epref{eq:GR} with $\hat \Sigma$ replaced by a negative imaginary
infinitesimal, $-\iu\hat 10^+$.
We can convince ourselves, that
\begin{align*}
\hat Q\, \hat g^\text{(R)}(E)^{-1} &= \hat Q \left[ \qty(E+\iu 0^+) \hat 1 - \hat H \right ] = \\
&= \phantom-\left[ \qty(E+\iu 0^+) \hat 1 + \hat H \right] \hat Q =\\
&= -\left[ \qty(-E-\iu 0^+) \hat 1 - \hat H \right ] \hat Q =\\
&= - \hat g^\text{(A)}(-E)^{-1} \hat Q,
\end{align*}
where, as before, we assumed SL symmetry in Model 2 and used \pref{eq:GA}.
Multiplication of the above equation from the right by $\hat g^\text{(R)}(E)$
and from the left by $\hat g^\text{(A)}(-E)$ and flipping signs
yields $\hat g^\text{(A)}(E)\hat Q = - \hat Q \hat g^\text{(R)}(-E)$. This is a re-formulation
of SL symmetry, Eqs.~(\ref{eq:com1},\ref{eq:ph}),
for isolated chains. We now define the SL symmetry of chains coupled
to reservoirs by the condition
\begin{equation}
\label{eq:QG}
\hat Q \GRof{E} =  - \GAof{-E} \hat Q.
\end{equation}
By repeating the calculation in the paragraph above with the full Green's function,
the condition \pref{eq:QG} implies
\begin{equation}
\label{eq:QS}
\hat Q \hat\Sigma({E}) =  - \hat\Sigma^\dagger({-E}) \hat Q.
\end{equation}
Using \pref{eq:SigmaDecomp}, 
\begin{subequations}
\label{eq:QDG}
\begin{align}
\hat Q \hat\Delta({E}) &=  - \hat\Delta({-E}) \hat Q\\
\hat Q \hat\Gamma({E}) &=  + \hat\Gamma({-E}) \hat Q.
\end{align}
\end{subequations}
Self-energy, representing the coupling to the leads, should fulfill these
conditions in order to preserve SL symmetry. The
components $\GaLof{E},\GaRof{E}$ inherit these conditions as well.
We remark that local coupling has not been assumed here [unlike 
\eg{} in \epref{eq:sigmas}].
Evidently, the
simplified cylindrical wide-band Ansatz \pref{eq:sigmas}, used in numerical calculations,
satisfies these conditions. Moreover, any constant contribution to
$\hat \Delta(E)$, that has the same structure as the SL-symmetric
end-group terms [\epref{eq:hpauli} with \pref{eq:ph}],
also fulfills \epref{eq:QDG}. Ultimately, every realistic lead is going to break
the SL symmetry. When the self-energy is obtained from a microscopic
description of the contacts (\eg{} \textit{ab-initio}),
the expressions put forth here can be used
to identify the SL-perturbing terms.

\subsection{Implications of sub-lattice symmetry for observables}
For both Model 1 and 2, the SL symmetry has the algebraic form of the anti-commutator,
cf. \epref{eq:com1} with \epref{eq:acomModel1}. The $\hat Q$ operator
takes on the role of $\hat P$ when we go from Model 1 to 2, compare \epref{eq:Pcoms} with
Eqs.~(\ref{eq:com2},\ref{eq:com3}). There is only one algebraic difference,
the presence of the minus sign in \pref{eq:Qprop} (unlike: $\hat P^{-1}=\hat P$).
The minus in \pref{eq:Qprop} disappears if one drops the factor i in the definition of $\hat Q$,
but we shall not do it.
We conclude that the consequences of SL symmetry are going to be identical for Model 1 and 2,
specifically, for the energy spectrum, helicality and angular momentum.
These consequences are shown in the following. We focus on Model 2 (oligoynes) for concreteness.

\subsubsection{\label{sec:qisol} Isolated chains - spectrum and helicality}
Let us have a Hamiltonian $\hat H$ of Model 2 with the eigenvalues and
eigenstates from
\[
\hat H \ket{\psi} = E \ket{\psi}.
\]
Then $\hat H\hat Q \ket\psi = - \hat Q\hat H\ket \psi = -E\hat Q\ket\psi$ if condition \pref{eq:ph} holds.
Hence, $\hat Q \ket{\psi}$ is a 'partner' eigenstate of $\hat H$ corresponding to eigen-energy $-E$.
These are the Coulson-Rushbrooke pairs, mentioned in the Introduction.
For instance, $\hat Q$ maps HOMO to LUMO
when the chain is half-filled with electrons. When the Fermi level coincides
with $E=0$, single-particle excitations (of this effectively non-interacting system) are
\textit{particle-hole symmetric}.

Moreover, for the expectation value of helicality in an arbitrary state $\ket{\varphi}$ it
holds that
\begin{equation}
\label{eq:hpsi}
\bra{\hat Q\varphi}  \hat h \ket{\hat Q\varphi} = - \bra{\varphi}\hat Q^\dagger \hat Q \hat h \ket{\varphi}
= -\bra{\varphi} \hat h \ket{\varphi},
\end{equation}
where Eqs.~(\ref{eq:com3},\ref{eq:Qprop}) have been
invoked. We conclude that the Hamiltonian eigenstates are mirror symmetric with respect to
energy and anti-symmetric with respect to their expectation value of helicality
when SL symmetry is present. For example, HOMO and LUMO have opposite winding sense,
and so do HOMO-1 and LUMO+1 (half filling).
These identities of the eigenstates and eigenvalues of Model 2 apply equally to
Model 1 (cumulenes). One only needs to replace $\hat Q$ for $\hat P$ in the derivations.

\paragraph*{Discussion of previous works on cumulenes.}
In Refs.~\cite{Hendon2013,Garner2019}, molecular
orbitals of various cumulenes were calculated \textit{ab-initio}. The helical HOMOs and LUMOs,
portrayed therein, have always opposite winding senses. The same was observed by Gunasekaran and Venkataraman, who
employed a H\"uckel model \cite{GunasekaranHelical}, equivalent to our Model 1. This phenomenon
can now be rationalized by the SL symmetry.

\subsubsection{\label{sec:qodd} {Transport: reciprocity of linear response coefficients}}
The SL symmetry \pref{eq:ph}, combined with time-reversal invariance, implies
certain identities of response functions, introduced in Sec.~\ref{sec:obs}.
These relations can be classified as Onsager reciprocity 
relations \cite{Onsager1, Onsager2, Jacquod2012}. \revision{}{We prove them 
within the Landauer formalism below. A discussion of the impact
of molecular vibrations is offered in Appendix~\ref{app:vibrations}.}

\paragraph*{Transmission.} 
To prove the consequences of the (anti-) commutators on the transmission
probability $\trans (E)$, we insert into the trace formula \pref{eq:te}
the identity $\hat 1 = -\hat Q\hat Q$ [see \epref{eq:Qprop}] and commute one
$\hat Q$ to the right, using Eqs.~(\ref{eq:QG},\ref{eq:QDG})
and cyclic invariance of the trace. Hence,
\begin{align*}
\trans (E) &= -
\text{Tr} \qty[\hat Q\hat Q
\GRof{E}\hat\Gamma_\text{L}(E)\GAof{E}\hat\Gamma_\text{R}(E)
] = \\
&=\phantom{-}
\text{Tr} \qty[\hat Q\GAof{-E}\hat\Gamma_\text{L}(-E)\hat Q
\GAof{E}\hat\Gamma_\text{R}(E)
] = \\
&=
-\text{Tr} \qty[\hat Q\GAof{-E}\hat\Gamma_\text{L}(-E)
\GRof{-E}\hat\Gamma_\text{R}(-E)\hat Q
] = \\
&= \phantom{-}
\text{Tr} \qty[\GAof{-E}\hat\Gamma_\text{L}(-E)
\GRof{-E}\hat\Gamma_\text{R}(-E)]
.
\end{align*}
Since the above formula looks like a transmission of some time-reversed
system, we insert there $\trev^2 = \hat 1$ under the trace and
proceed commuting as before,
\begin{align*}
\trans (E) &= \text{Tr} \qty[\trev \trev \GAof{-E}\hat\Gamma_\text{L}(-E)
\GRof{-E}\hat\Gamma_\text{R}(-E)] = \\
 &= \text{Tr} \qty[\trev \GRof{-E}\hat\Gamma_\text{L}(-E)\trev 
\GRof{-E}\hat\Gamma_\text{R}(-E)] = \\
 &= \text{Tr} \qty[\trev \GRof{-E}\hat\Gamma_\text{L} (-E)
\GAof{-E}\hat\Gamma_\text{R}(-E)\trev].
\end{align*}
Eqs.~(\ref{eq:trevg},\ref{eq:trevGamma}) have been used. 
Above, the right $\trev$ 
\revision{has nothing left to conjugate}{operates on real wavefunctions}
and the left one
can be factored before the trace (as a conjugation),
which is real.
We have achieved a standard result:
\begin{equation}
\label{eq:tes}
\trans (E) = \trans (-E).
\end{equation}
Indeed, it is known that transmission of 
a time-reversal and SL -- symmetric
system is even; see \eg{} Ref.~\cite{Jacquod2012}, where the statement is derived within
a family of \textit{Onsager} reciprocity theorems.

\paragraph*{{Helicality and angular momentum.}}
Now we are in a place to prove the identities
for the linear response coefficients of helicality and angular
momentum, Eqs.~(\ref{eq:dhde},\ref{eq:dlz0}). These response
coefficients are odd functions of the Fermi energy $E=\EF$, namely,
\begin{subequations}
\begin{align}
\label{eq:h0es}
\pdv{h(E,0)}{E} &= -\pdv{h(-E,0)}{E}\\
\label{eq:lzes}
\evaluated{\pdv{L_z(E,V)}{V}}_{V=0} &= -\evaluated{\pdv{L_z(-E,V)}{V}}_{V=0}.
\end{align}
\label{eq:coeffe}
\end{subequations}
To prove the latter relation we start with the respective trace expression
\begin{align*}
\text{Tr} \qty{ \hat L_z\, \GRof{E} \qty[\GaLof{E} - \GaRof{E}] \GAof{E} }
\end{align*}
and insert $\hat 1= -\trev^2\hat Q^2$, in full analogy to the treatment of the
transmission. Additional minus sign appears:
$\hat L_z$ commutes with $\hat Q$ but anti-commutes with $\trev$.

Finally, from
\[
\hat Q \qty[ \GRof{E} -\GAof{E}]=  
\qty[\GRof{-E} - \GAof{-E}] \hat Q
\]
we infer that 
\[
\hat Q\hat A(E) = \hat A(-E) \hat Q
\]
[see the definition of $\hat A(E)$ in \epref{eq:spectralf}].
Combining with \pref{eq:com3} we have $\text{Tr} \qty[ \hat Q\hat Q\hat h \hat A(E)]
=-\text{Tr} \qty[ \hat Q \hat h  \hat A(-E)\hat Q]$,
henceforth \epref{eq:h0es} is proven. This identity does not refer to a non-equilibrium
and the derivation does not involve time-reversal, it is therefore not
an Onsager reciprocity.

\section{Helicality in perturbation theory \label{sec:pert}}

\begin{figure}
\includegraphics[width=0.45\columnwidth]{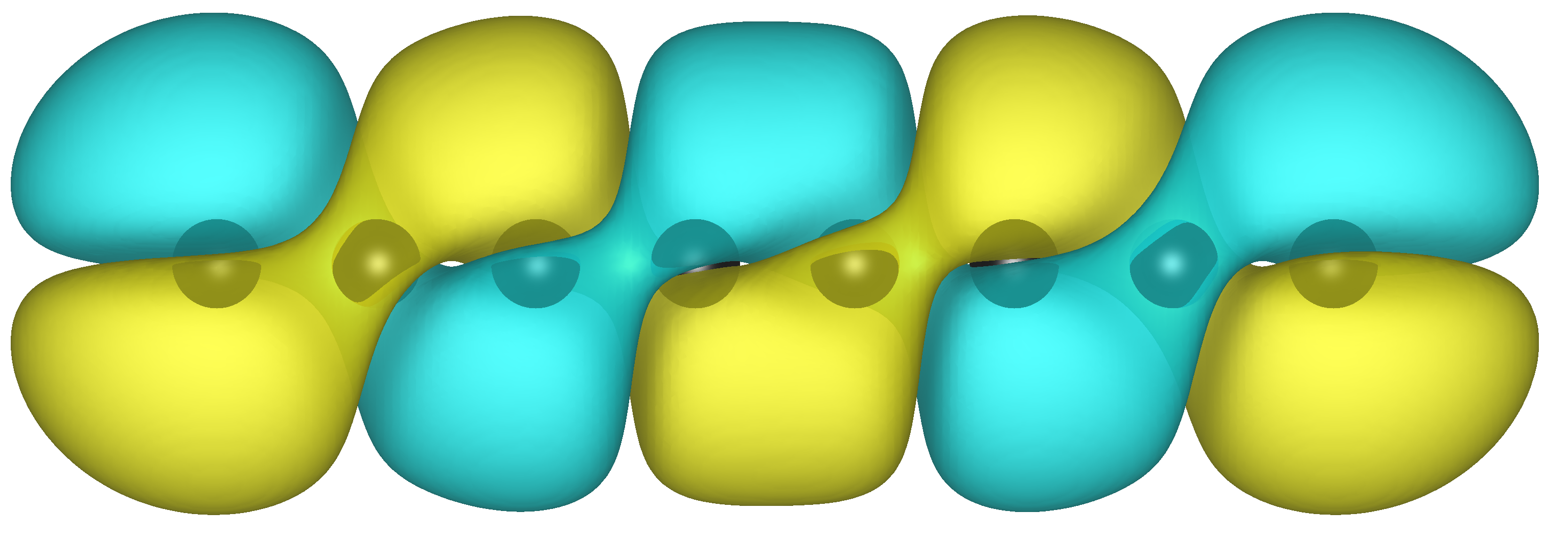}
\includegraphics[width=0.45\columnwidth]{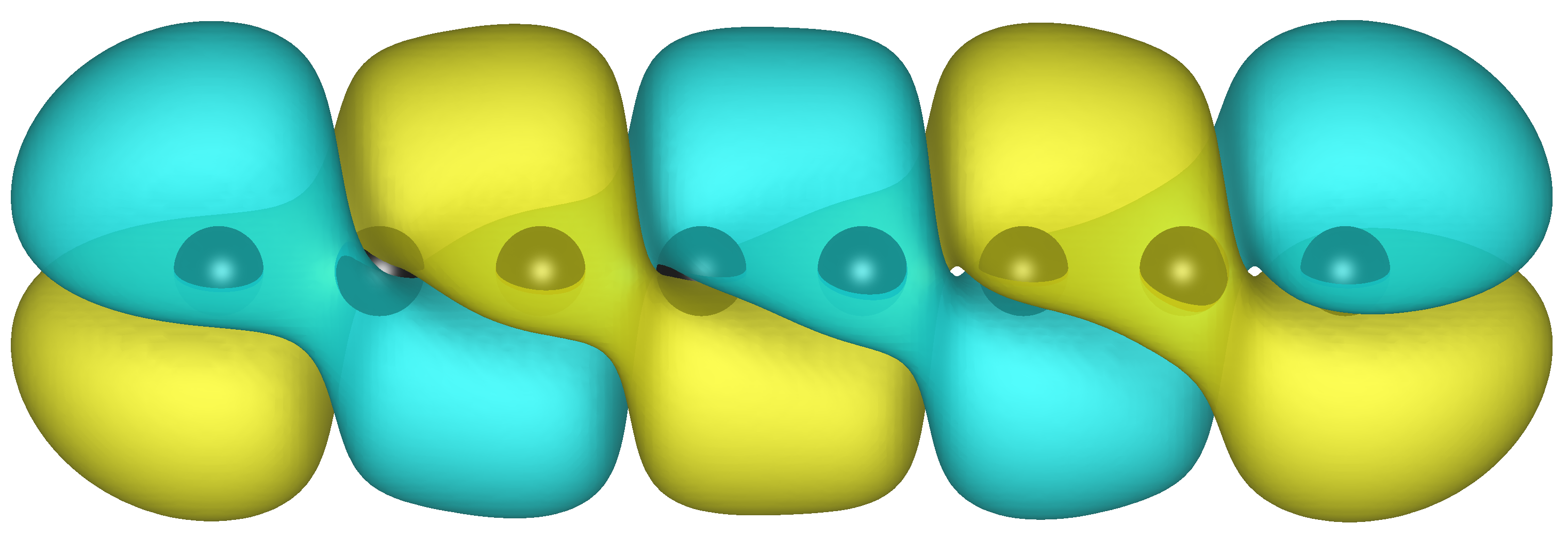}
\includegraphics[width=0.45\columnwidth]{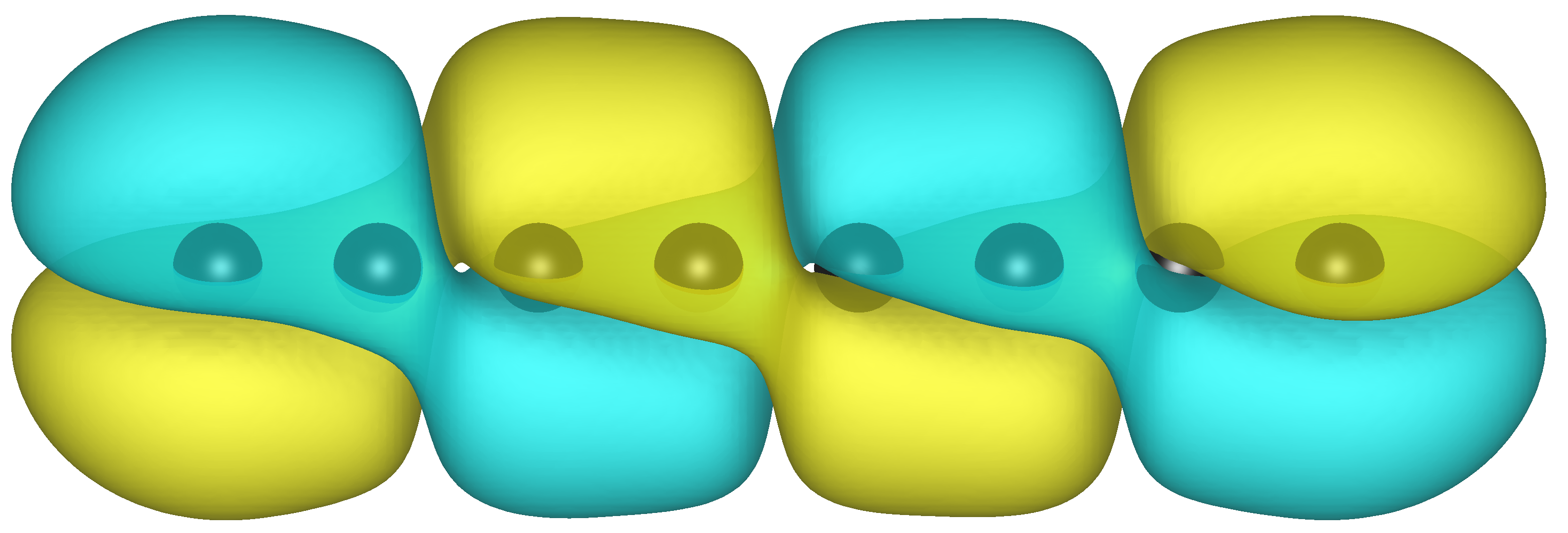}
\includegraphics[width=0.45\columnwidth]{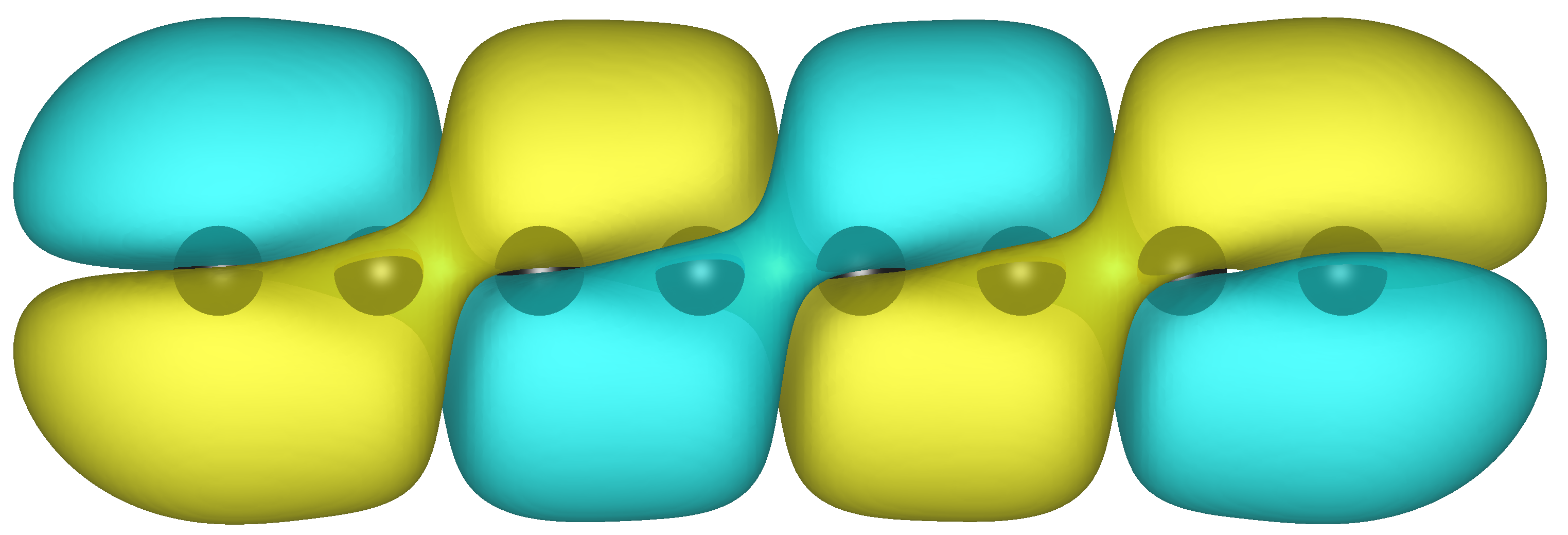}
\caption{The frontier orbitals of carbon chains ($N =8$)
with the end-groups parameterized with $\epsilon=0.5t$, $\thetat = 45^\circ$:
LUMO+1, LUMO (upper row), and HOMO--1, HOMO(lower row). The winding sense of
the \revision{helicality}{orbitals} is alternating with increasing orbital energy.
Images generated with VESTA\cite{VESTA}.
}
\label{perturbationOrbitals}
\end{figure}

To motivate the subsequent perturbative analysis, we
plot
in \figref{perturbationOrbitals} few orbitals with energies around the center
of the band for
$\thetat=45^\circ$, $N=8$,$\epsilon=0.5t$ (Model 2).
The SL symmetry is only broken weakly (see \appref{app:sublattice}) and the
band center is located between HOMO (highest occupied molecular orbital) and
LUMO (lowest unoccupied molecular orbitals). The portrayed molecular
orbitals display winding directions consistent with \epref{eq:hpsi}.
Furthermore,
the sign of helicality alternates with increasing orbital energy,
which we inspect here analytically.
For $|\epsilon| \ll t$, we can employ
perturbation theory in $ \hat H_\text{pert} = \hat H_1 + \hat H_N$. The $\hat H_0$
is a reference Hamiltonian, representing unsubstituted oligoynes.
Because $\hat H_0$ has the 
p$_x$, p$_y$ subsystems decoupled, the eigenstates are doubly degenerate.
We list the unperturbed 
eigenstates and eigenenergies in \appref{app:perturbationApplication}.

The first-order energy corrections are
\begin{align}
\label{eq:e1}
    E ^{(1)} _ {\mu \pm} = \frac{2 \epsilon}{N+1} \sin ^ 2 \left ( \frac{\mu \pi}{N+1} \right ) \left ( 1 \pm \cos \left (\thetat \right )\right )
\end{align}
where $\mu \in \{1, 2, 3, ..., N\}$ is the quantum number of the
unperturbed eigenstates 
and $\pm$ label transverse chain modes
(see \appref{app:perturbation} or \cite{MarekThesis} for details).

The expectation value of helicality, to first order, results in
\begin{align}
\label{eq:h1}
    h ^{(1)} _ {\mu,\pm} = &
    \pm \frac{4 \epsilon}{t(N+1) ^ 2} \sin(\thetat)  \nonumber \\
    &\times \sum _{\substack{\nu \\ \mu + \nu= \text{ odd}}} \frac{\sin ^ 2 \left ( \frac{\mu \pi}{N + 1}\right )
    \sin ^ 2(\frac{\nu \pi}{N+1})}{\left ( \cos \left (\frac{\mu \pi}{N+1} \right ) -
    \cos\left ( \frac{\nu \pi}{N + 1} \right )\right ) ^ 2}
\end{align}
(see \appref{app:perturbation} or \cite{MarekThesis} for derivation).
The last formula shows that helicality can be broken down into 
several factors: The term $\sin(\thetat)$ shows that the $\pi$
orbital is twisted due to the boundary terms. Helicality vanishes for
angles $\thetat = 0,\pm \pi$. In our model, these special angles imply a
planar (D$_{2\mathrm h}$) symmetry, which excludes helical eigenstates, as discussed
by Hendon and coworkers \cite{Hendon2013}. The terms included after the summation
in \pref{eq:h1}
account for the influence of the variation of the longitudinal wavefunction phase onto the precession
of $\bm P_n$. 
Importantly, comparing Eqs.~(\ref{eq:e1},\ref{eq:h1}),
the originally degenerate p$_x$, p$_y$ doublets 
always split into pairs of states with
opposite helicalities. The only exceptions are
$\thetat = 0,\pm \pi$ (no helical orbitals). Additionally, for
$\thetat = \pm \half\pi$ the pair of
orbitals is degenerate; helicality can be removed by an orthogonal transformation.
The alternating helicality was observed already
by Gunasekaran and Venkataraman\cite{GunasekaranHelical}
for $\epsilon=\infty$; here we show that it reproduces in the perturbative regime as well
and point out its physical consequence (for angular momentum) in the next section.
\figref{perturbationGraph}
shows that the perturbation theory works qualitatively well for both
energy and helicality even when $\epsilon \approx t$.

\begin{figure}[ht]
    \includegraphics[width=\columnwidth]{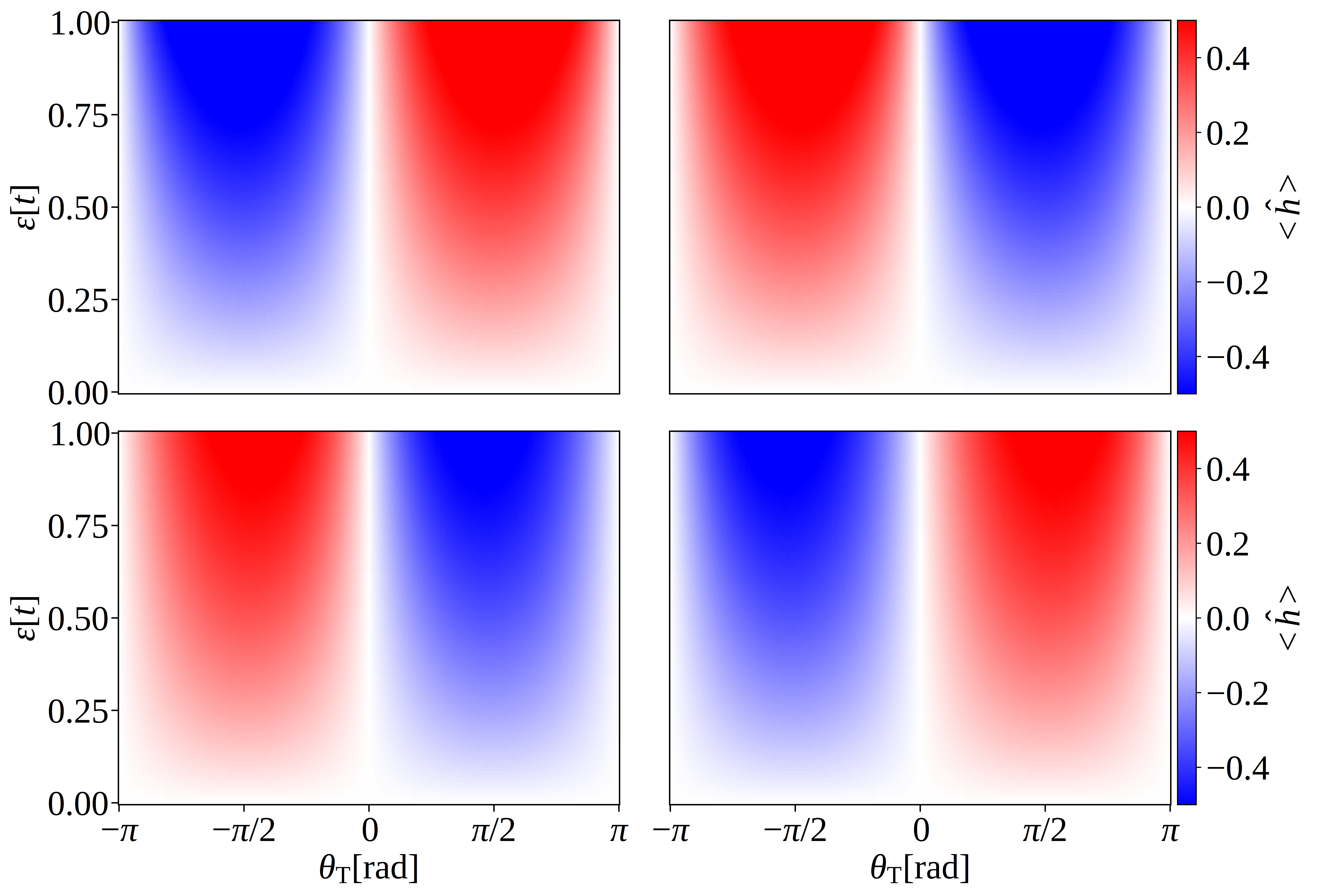}
    \caption{The helicality $\expval{\hat h}$ as a function of $\thetat$ and $\epsilon$:
     HOMO--1, HOMO (bottom row), and LUMO+1, LUMO (top row) for $N=8$.}
    \label{perturbationColormap}
\end{figure}

To inspect the alternation further, we show the helicality beyond perturbative 
regime from numerical diagonalization in \figref{perturbationColormap},
showing a monotonous dependence of $\expval{\hat h}$ on $\epsilon$
which is free from sign changes.
Naturally, a large $\epsilon$ can lead to level crossings, which
break the alternation pattern. Such crossings occur first near
the bottom and top of the band. The
states near the center of the band will still preserve alternation
for $\epsilon$ up to $\approx t$.
In view of the Ref.~\cite{GunasekaranHelical}, the alternation
re-establishes in the entire spectrum for large $\epsilon\to\infty$, representing
cumulenes (Model 2 goes to Model 1).
\revision{}{Finally, we remark that the chain characteristics discussed here analytically 
and numerically are robust against moderately-strong substitutional disorder,
we refer to \appref{app:dis} for details.}

\begin{figure*}
\includegraphics[width=.8\textwidth]{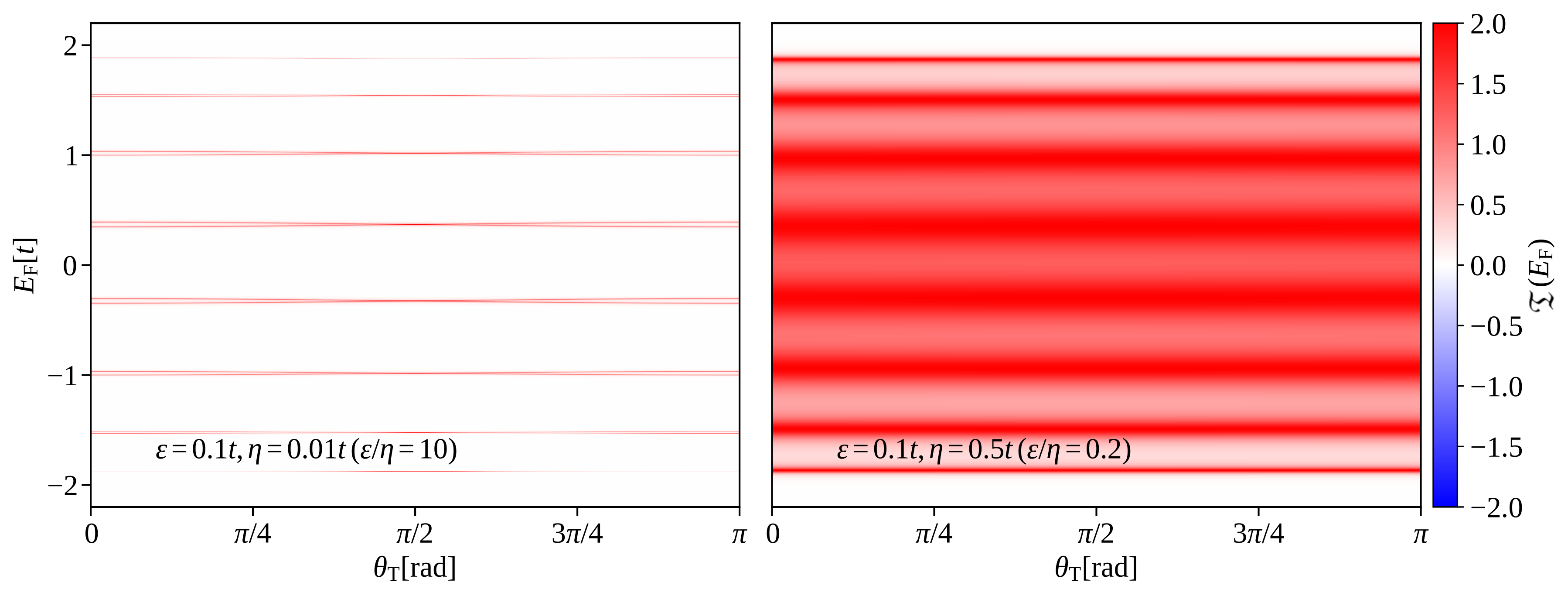}\\
\includegraphics[width=.8\textwidth]{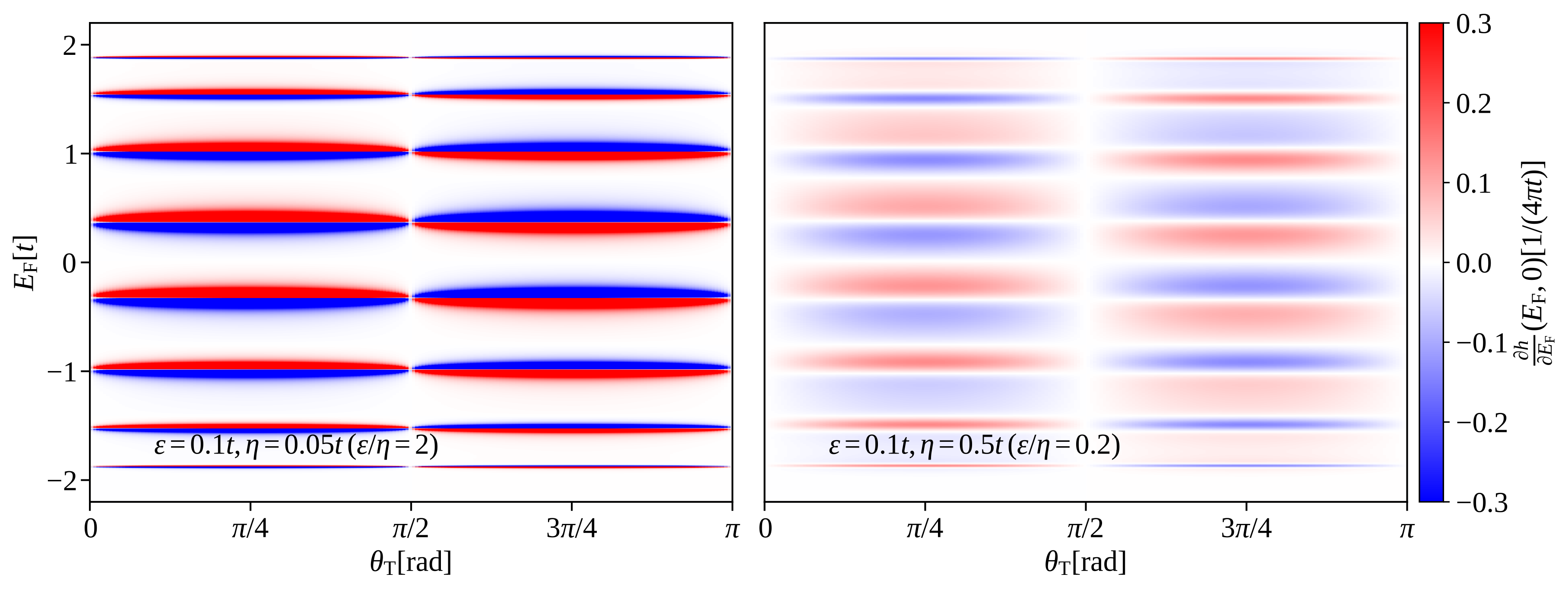}\\
\includegraphics[width=.8\textwidth]{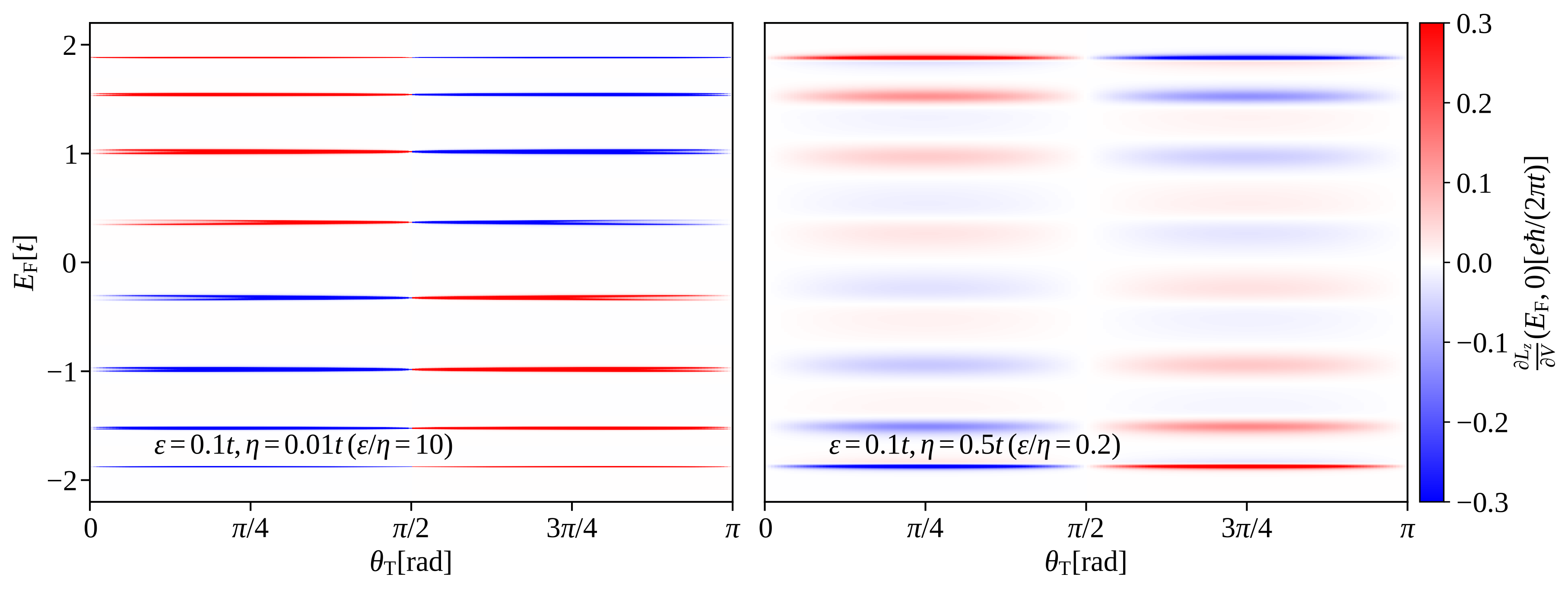}
\caption{Color maps of various observables for the chain with $N=8$
sites coupled to reservoirs with $\eta =0.01t$ (left column) and
$\eta=0.5t$ (right column) for twist angles $\thetat>0$, with $\epsilon=0.1t$.
Top row: transmission function $\mathfrak T(E)$, second row: equilibrium helicality
$\qty(\partial h/\partial \EF)$,
third row: angular momentum response $\qty(\partial L_z/\partial V)$.
\label{fig:triple_graph}}
\end{figure*}

\section{Chains coupled to reservoirs: Numerical results \label{sec:num}}
Coupling the carbon chains to reservoirs under bias
results in circular currents that wind around the axis.
It is natural to expect that the associated circular currents produce
an orbital angular momentum. Here we explore in detail the relation
between the latter and helicality.

\subsection{Angular dependence}
To relate the emergence of angular momentum to orbital helicality, we
show various spectral densities of embedded chains in \figref{fig:triple_graph}.
The program and generated data are available in \cite{NumericsZenodo}.

The first row shows the transmission function [\epref{eq:te}]. 
For $\eta =0.01t$ (left column), the quasi-degenerate
doublets are clearly visible.
Their angular dependence is consistent with the behavior of eigenenergies of isolated chains,
\figref{perturbationGraph} and \epref{eq:e1}.
In the strong coupling regime, $\eta =0.5t$ (right column), the resonances
are much wider, as expected. For the subsequent analysis the following distinction
becomes useful: In the weak coupling,  only the states within the doublet can spectrally
overlap, while in the strong coupling, states from the nearby doublets may admix.

The second row presents the equilibrium coefficient,
$\partial h(\EF, V=0)/\partial \EF$.
The plots can be understood by recalling the analytic result \pref{eq:e1},
or \figref{perturbationGraph}, for the splitting of p$_{x,y}$ doublets
into pairs of opposite helicality. For both strong and weak couplings, \revision{the}{} helicality
vanishes trivially at $\thetat=0,\pi$ and at  $\thetat=\frac \pi 2$, because
this is the degeneracy point -- the transmission resonances overlap completely
and no net helicality is generated. The sectors
$\thetat\in\langle 0, \frac\pi 2\rangle$ and $\langle \frac \pi 2, \pi\rangle$ differ only by the sign,
because at the degeneracy point the two quasi-denegenerate states interchange.
Helicality also vanishes for $\EF\approx 0$, when the Fermi level is in the middle of the
HOMO-LUMO gap (``half filling'').
At weak coupling, when only the quasi-degenerate pairs 
have spectral overlap, $\partial h(\EF, 0)/\partial \EF$ is vanishingly small for most $\EF$, 
except near transmission resonances. This is because the pairs have
opposite helicality and their integrated effect on the equilibrium helicality cancels.
Also in the strong coupling
regime do we find sign alternation across the energies: for fixed $\thetat$,
there is a sign change (node) at the original doublet energies and another
node in the larger gaps between doublets (off-resonant energies).
We notice that for $0<\thetat<\frac\pi 2$ and $\EF<0$ the latter node is
pushed down in energy from the midpoint energy of the larger gap. In other words,
in the quadrant $0<\thetat<\frac\pi 2$ and $\EF<0$, the negative $h(\EF,0)$
occupies more area.

The bottom row of \figref{fig:triple_graph} presents the angular momentum.
Crucially, for $\thetat = 0, \pi/2, \pi$ helicality vanishes
(`linear polarization'); consequently, angular momentum vanishes as well.
The most important observation is, that
 $\partial L_z(\EF, V=0)/\partial V$ is an odd function with respect to the middle of the HOMO-LUMO gap
($\EF =0$). The details of the
angular momentum response differ at strong and weak couplings, respectively.
For $\eta =0.5t$ the angular momentum response follows closely the helicality response; in particular,
its signs and nodes. Loosely speaking, the angular momentum inherits the alternating behavior (as a function of $\EF$)
from helicality. In weak coupling, for a fixed $\thetat$,
the behavior of $\partial L_z(\EF,V=0)/\partial V$ is markedly different from $\partial h(\EF, V=0)/\partial \EF$.
For $\eta =0.01t$, the pairs of perturbatively split resonances (doublets) manifest as 
doublets in $\partial L_z(\EF, V=0)/\partial V$, however, with equal sign. For a fixed
$\thetat$, the sign change occurs markedly at $\EF$.
The change of the $\partial L_z(\EF,V=0)/\partial V$ peak structure between
the two regimes (strong and weak coupling) is explored in the next section.

These numerical results exemplify the preceding analytical
analysis: (1) helicality alternates with energy, result \epref{eq:h1}.
We add to this\revision{ the}{} observation the following remark:
When time-reversal is broken by the electric current, the
alternating helicality translates into the alternation of angular
momentum, with $\EF$.
(2) The symmetry relations \pref{eq:coeffe} and \pref{eq:tes}
manifest in \figref{fig:triple_graph}. The SL symmetry,
although approximate, is a powerful concept to rationalize the
energy dependence of the observables. We inspect the departure
of SL symmetry in more detail in the subsequent sections.

\subsection{From weak to strong coupling}
\begin{figure}
\includegraphics[width=0.95\columnwidth]{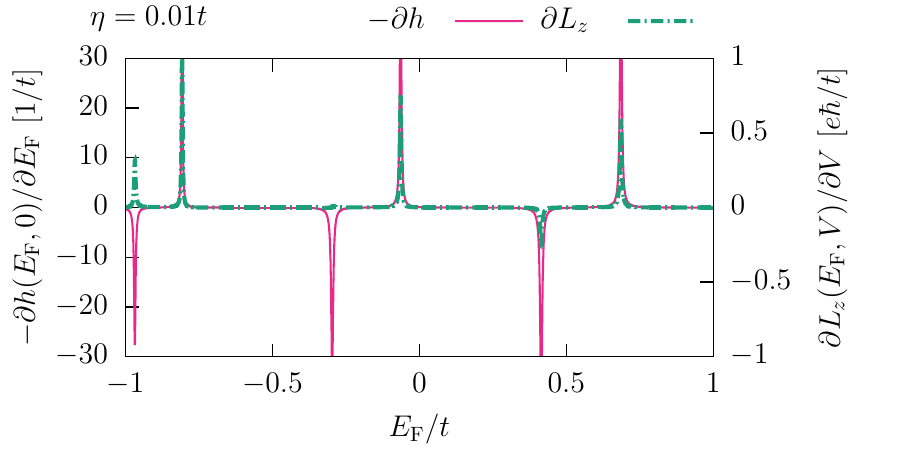}
\includegraphics[width=0.95\columnwidth]{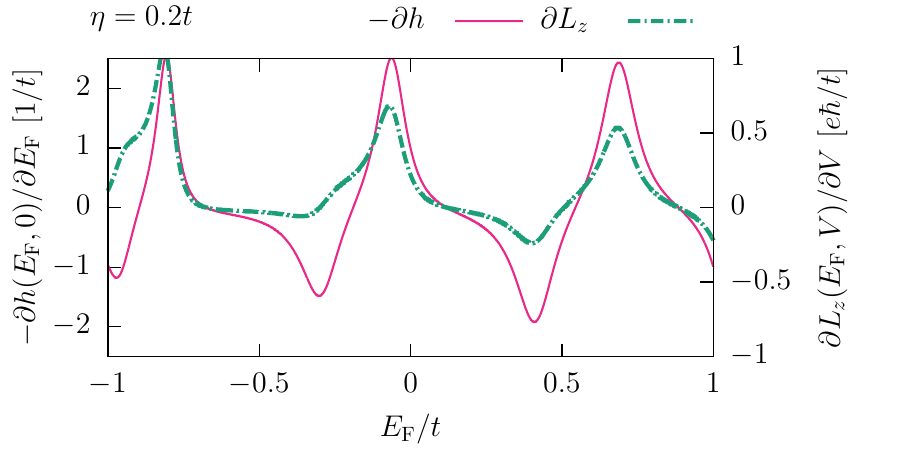}
\includegraphics[width=0.95\columnwidth]{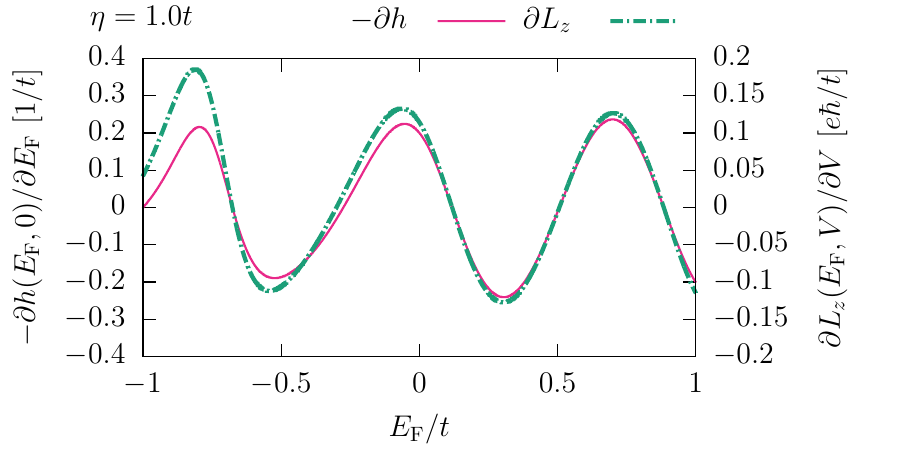}
\caption{\label{fig:couplingDep}
Comparison of helicality, $(\partial h/\partial \EF)$, (solid magenta lines) 
with angular momentum, $(\partial L_z/\partial V)$, (dot-dashed green)
 responses for couplings $\eta=0.01t$ (top),
$0.2t$ (middle) and $\eta = 1t$ (bottom). Notice that helicality is multiplied by a minus sign.
The parameters are $N=8, \epsilon = 0.1t, \thetat = \frac\pi 4$.}
\end{figure}

To reveal the interdependence of helicality and angular momentum, we 
fix here the twist angle to $\thetat=\frac{\pi}4$ and plot
the linear responses in \figref{fig:couplingDep} for various coupling strengths.

We first notice that $\partial L_z(\EF,V=0)/\partial V$ and
$\partial h(\EF, V=0)/\partial \EF$
are mostly of opposite signs. This can be rationalized using the
continuum form, \epref{eq:important}. Since the $z$ coordinate increases from the
left end-group towards the right one, and particles flow from left to right when
$V$ is positive, the linear momentum, on average, must be positive. The minus
sign in \figref{fig:couplingDep} thus reflects the minus sign in \epref{eq:important}.
Weak SL symmetry breaking manifests here by the $\EF$ dependencies being 'almost odd'
with respect to energy $E\approx 0.1t$.

The noteworthy observation is that the 'negative' helicality and angular momentum follow
closely each other, when the coupling is strong. Since 
$\partial h(\EF, V=0)/\partial \EF$ can be understood from the helicality of chains
in isolation (helicality of molecular orbitals), 
helicality becomes a proxy for the resulting angular momentum upon the current flow,
in non-equilibrium. This correspondence seems to \revision{weaker}{weaken} for weak coupling,
but still seems to hold near $E=0$.

\subsection{Angular momentum beyond linear response\label{sec:nonlin}}
\epref{eq:lzes} states that when particle-hole symmetry is present, \ie{}
there is SL symmetry and the Fermi level is in the band center, $\EF=0$,
angular momentum is zero in linear response in the voltage bias. Remarkably,
the sign change of $\partial L_z(\EF, V=0)/\partial V$ at $\EF =0$ 
has a peculiar physical manifestation beyond linear response.
For definiteness, we assume that the bias drops entirely on the left
lead, $\mu_\text L = \EF + eV$ while $\mu_\text R = \EF$. This description is used 
to model scanning-probes with molecules
on surfaces; the molecule equilibrates with the surface (here the right lead).
To express the non-linear response of $L_z(\EF, V)$, we
invoke Eqs.~(\ref{eq:dodv},\ref{eq:ononlin}) and recognize the relation
\begin{equation}
\label{eq:lzstm}
L_z(\EF, V) = \frac 1e \int_{\EF}^{\EF + eV}
\pdv{L_z(E,V=0)}{V} \dd E.
\end{equation}
The latter is even in $V$. Angular momentum does not reverse sign when the current
flow reverses.

\begin{figure}
    \centering
    \includegraphics[width=\columnwidth]{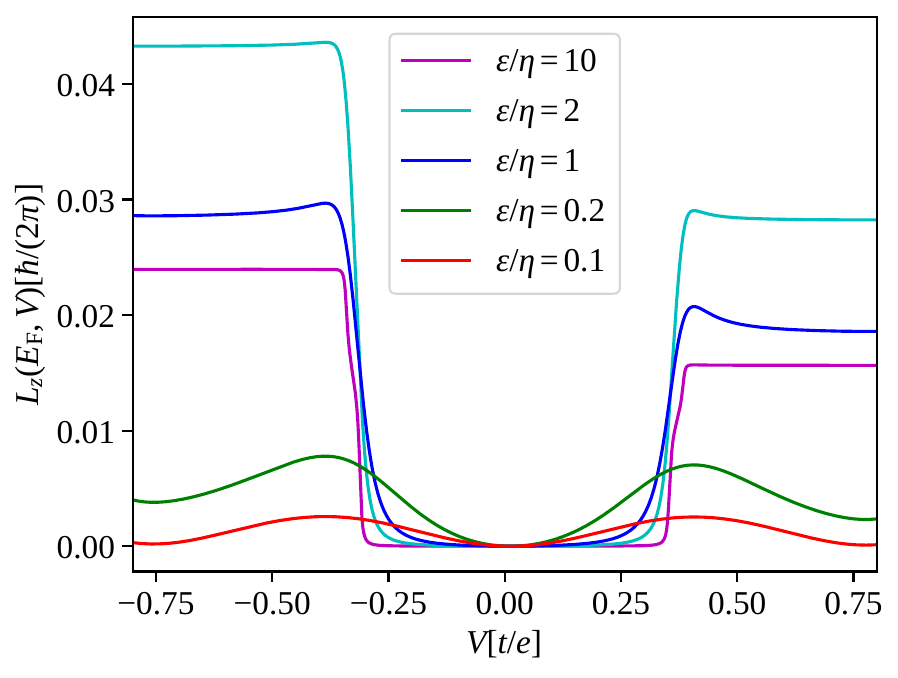}
    \caption{The expectation value of the angular momentum as a function of $V$ for $\EF=0$.
    Parameters: $\thetat \approx -44.66^\circ, N=8, \epsilon=0.1t$.}
    \label{biasSymmetry}
\end{figure}

The connection to helicality offers an intuitive understanding of this result.
When the bias is positive, the tunneling current 'probes'
molecular orbitals that are different when the bias is 
reversed, \eg{} these can be the HOMO and the LUMO.
Due to SL symmetry, these orbitals have opposite helicality.
If current runs in the same direction,
 intuitively, these orbitals carry opposite angular momentum as well.
 A classical picture of a particle traveling along screws is invoked. However, the sign
of the current is reversed with $V$ and this implies that $L_z(\EF=0, V)$ does not
change sign. This phenomenology holds true not only for Model 2 (oligoynes), but also Model
1 (cumulenes), as established in Sec.~\ref{sec:slcumu}.
\figref{biasSymmetry} showcases these ideas numerically for different coupling
strengths.

The figure reveals another remarkable feature of the voltage dependence
of the induced angular momentum. When $|eV|$ exceeds the transmission resonance energy
$\sim 0.35t$ of the frontier orbital, $L_z(\EF,V)$ decreases.
This decrease is stronger at strong coupling. The observation
can be rationalized in virtue of the discussion in the preceding
section and \figref{fig:couplingDep}. Since orbital helicality
alternates, raising the voltage beyond the resonance energy
admixes the subsequent orbital and the latter contributes
to the integral \pref{eq:lzstm}
with angular momentum of an opposite sign. An experimental-based
study \cite{SkolautThesis,ExperimentalPaper} also reported a decrease of rotation for large
bias, despite an increase of the current.

We end by noting that the action of SL symmetry-breaking terms
[$\propto \s{0}$ in \pref{eq:hpauli}]
is evident in \figref{biasSymmetry}. The difference in the large $|V|$ values of
$L_z(\EF,\pm V)$ is $\approx 20\%$ \revision{}{of average value $(L_z(\EF, V) + L_z(\EF, -V))/2$}. Notice that the SL-breaking terms have the
same energy $\epsilon$ as the helicality and angular-momentum -- inducing terms,
in the chosen parameterization.
Yet, the SL breaking does not change the qualitative picture.

\section{Discussion and outlook}
Previous sections predict that 
there is a finite electronic orbital angular momentum (EOAM)
in a situation with a steady-state current through the carbon chain (axle).
The EOAM is here linked to helical orbitals, but this
is not a necessary condition -- it is natural to expect the EOAM
in molecular junctions with axial chirality.
Here we discuss the implications of the steady-state EOAM on the
dynamics of the molecular rotor.

\paragraph*{Review of a minimal model.} We
invoke here a minimal model of the current-induced molecular rotation
introduced by Koryt\'ar and Evers in Ref.~\cite{KorytarMomentum}.
This minimal model relies on angular momentum conservation,
which is respected by quantum effects as well.

\begin{figure}
\includegraphics[width=.9\columnwidth]{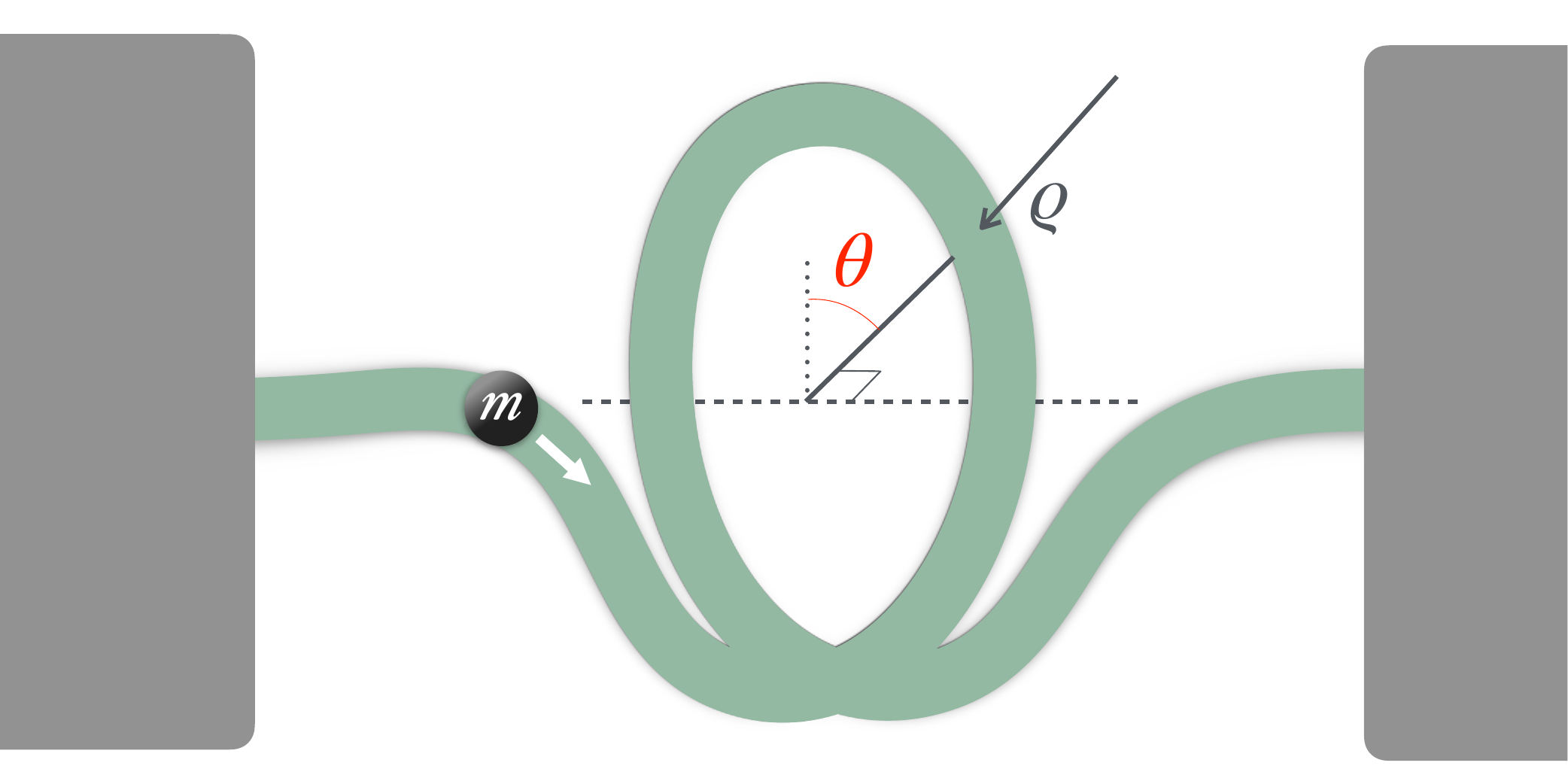}
\caption{\label{fig:tube}A minimal model of current-induced rotation
with a molecular junction represented by a helical tube.
Here, the helix emerges from one lead, it revolves approximately $N=1$ times
and connects to the other lead. With respect to the central axis
(dashed line), the
initial and final radius of the trajectory is zero, and the radius
attains its maximum $\varrho$. A particle
of mass $m$ can move freely through the tube. The particle does
not carry angular momentum, except when it is in the central helical segment.
The tube can rotate freely around the axis -- its orientation $\theta(t)$
is a dynamic variable.}
\end{figure}
Let us recall the essence of this model for the convenience of the reader.
Consider source and drain connected by a narrow helical tube (representing
the molecular junction) of the form shown in \figref{fig:tube}. 
Next, a particle (electron) can move freely along the tube,
being constrained by the walls of the tube, only.
A single scattering event, with the electron being
transported from the source to the drain inertially, without
acceleration from an external field, is envisioned.
The rotor is still before this process. The electron
acquires an EOAM in the tube, and this is compensated by the angular momentum
of the tube, $\imo \dot{\theta}$, where $\imo$ is the
rotor's moment of inertia.
Because the particle's angular momentum is zero in the source
and drain, there is no angular momentum of the rotor before and after the
scattering. Yet, the rotor is left turned by an angle
after the incident. The angle is
\begin{equation}
\label{eq:deltaThetat}
\Delta \theta = \pm \frac{2m \pi \expval{\varrho^2} N}{\imo },
\end{equation}
where $m$ is the particle (electron) mass, $N$ is approximately the number of turns of the helix and 
$\expval{\varrho^2}$ is the average squared radius of the helix.
The right-hand side does not depend on dynamical variables,
such as scattering time and the EOAM. It only depends
on the geometry and the mass ratio. We emphasize that the sign is determined
by the chirality -- it is opposite to the sign of EOAM
during the transport. There is a torque when the electron
enters the helix -- for without it there would be no turn --
and an opposite torque that stops the rotation
when the electron leaves the helix.
The mechanism induces rotation even if there is a potential barrier.

\paragraph*{Application to chains with helical orbitals.}
We propose here that several outcomes of the minimal model
apply to the carbon axles, connecting rotor and stator moieties
(as shown in \figref{f1}), and steady-state EOAM.
If the rotation is free (the potential barrier is low),
a constant angular velocity $\dot\theta$
is expected with steady-state current. This is because of a current flow:
particles entering and leaving the carbon axle boost the $\theta$.

An order-of-magnitude estimate of the boost $\Delta \theta$ per electron can be 
estimated by setting $\imo \approx M\expval{R^2}$,
with the averaged radius of the rotor $R$:
\begin{equation}
\Delta \theta \approx 10 \frac mM \frac{\expval{\varrho^2}}{\expval{R^2}}.
\end{equation}
If $m/M \approx 10^{-5}$ (for a few atoms) and the ratio of variances is $10^{-2}$, $\Delta \theta
\approx 10^{-6}$ rad. We have assumed $2\pi N \approx 10$ here.
A steady-state current scenario involves many
electrons passing from source to drain. If the electrons are assumed
independent, the current of 1 nA would induce full rotation in $\approx 10^{-4}$ sec.
The value of this estimate is that it provides an upper bound,
because, \eg{}, other tunneling pathways and inelastic effects have been ignored.

To study the functionality of helical orbitals for
molecular motors for specific applications, a more
detailed description, including \eg{} dissipation
and current-induced noise, is needed.
More realistic models involve
potential barriers \cite{KorytarMomentum} and semi-classical dynamics
\cite{Bode2011,Elste2008,Lu2012,Guttesen2025} that can involve ab-initio details \cite{Tao2015}
for specific molecular designs.

\paragraph*{Structural vs. orbital chirality.}
The direction of rotation is determined by the sign of
the EOAM, which in turn is determined by orbital helicality.
The direction of rotation can be independent on the sign of the current,
in view of Sec.~\ref{sec:nonlin}. Such a molecular motor
would operate as a galvano-mechanical rectifier.
\revision{}{This behavior can be contrasted with helix-shaped molecular
wires, \eg{} DNA, or chiral carbon nanotubes: there the OAM with respect
to the molecular axis must
change sign with the current reversal. This is dictated by the \textit{structural
chirality}. Carbynes and cumulenes own helical orbitals whose
helicality changes as a function of energy, allowing the galvano-mechanical rectification.}

\paragraph*{\revision{}{Effects of spin-orbit coupling (SOC).}}
\revision{}{While the SOC is weak in linear carbon chains, we would like to point out
that interesting behavior can happen when one end-group contains
a heavier element with sizable SOC (Si, Pt, see
Ref.~\cite{Gleiter2010} for more examples). When bias voltage is applied, this element
can translate the OAM of the carbon chain into a spin current. This opens a possibility
of chirality-induced spin-selective phenomena \cite{Korytar2024,Ruitenbeek2023} with unusual behavior under
current reversal.}

\section{Conclusions}
We have shown that carbon chains with helical orbitals lead to
finite expectation values of electronic angular momentum, when these chains
are coupled to reservoirs as molecular junctions. To make this relation more precise,
we have quantified helicality
of molecular orbitals by introducing a hermitian operator. For small twists and long chains,
the expectation value of the operator equals an average orbital twist angle per bond.

We have elaborated several implications of sub-lattice (SL) symmetry for helicality
and angular momentum. For the specific case of oligoynes, the SL symmetry
is hidden in the sense that it is not straight-forwardly 
identified from the Hamiltonian of the H\"uckel theory.
We have shown that molecular orbitals, whose energies are located symmetrically
around band center, have opposite helicality.

For chains coupled to leads with bias voltage $V$, the angular
momentum response coefficient has a node at the band center, as a consequence of
Onsager reciprocity. In the non-linear regime,
we show that this behavior can lead to the angular momentum not changing sign
with the current reversal.
We propose an application as a molecular
motor with a rectifying electromechanical response.
\revision{}{In this context we remark
that circulating currents also generate a magnetic field and its detection
was suggested earlier in Ref.~\cite{Walz2015}.}

Finally, we have clarified the
connection between helical orbital shape, brought about and discussed in chemistry,
and the helicity observable from the relativistic Dirac theory. In the
continuum limit, helicity and helicality (defined here) are identical, up to a natural constant.

\section*{Acknowledgments}
We thank Jan Wilhelm for helpful discussions in the early stage of the project.
We are indebted to Lukas Gerhard for help with writing of the manuscript.
We acknowledge the support of Czech Science Foundation (GAČR) through grant 22-22419S, 
the support of Charles University through GAUK (ID: 366222) fellowship and funding by the German Research Foundation (DFG) as part of the German Excellence Strategy - EXC3112/1 - 533767171 (Center for Chiral Electronics) and CRC 1277
(project number 314695032, subproject A03), and RTG 2905 (project number 502572516). 

\appendix

\section{Some properties of the helicality operator \label{app:operators}}
We elaborate on certain aspects of \revision{the}{} helicality defined in this work.
We give an alternative geometric picture in \appref{app:averagePitchDensityName}.
Then, in \appref{app:previousWork} we evaluate helicality for cumulene
eigenstates given elsewhere.
We derive the commutator with the Hamiltonian in \appref{app:averagePitchDensityCommutator}
and finally diagonalize the helicality
operator in \appref{app:eigenvectors}.

\subsection{Geometric view of helicality with a classical particle on a helix \label{app:averagePitchDensityName}}

Suppose we have a helix, such that if we travel
distance $2 \pi {P}$ along, we complete one turn exactly. Therefore,
the change of a polar angle of a point on the helix is
$\phi = \frac{z}{{P}}$ where $z$ is the
length of the arc described by the point. The rate
of change of the polar angle per distance is therefore
\begin{align}
    \frac{\partial \phi}{\partial z} = \frac{1}{{P}}.
\end{align}

We can apply this idea to the precessing $\bm P_n$ vectors from Sec.~\ref{sec:hop}.
There, the nodal plane of the $p$-orbitals may not be turning uniformly and therefore we choose
to focus on the average rate of rotation instead.
Furthermore, we investigate the sine of the polar angle,
as defined in \figref{PolarisationVectors}. 
We associate the
rate of change $\frac{1}{\tilde{{P}}}$ (per bond) with the following expression
\begin{align}
    \frac{1}{\tilde{{P}}} := \frac{1}{N-1} \sum _ {n=1} ^ {N-1}
    \frac{\bm P _ n \times \bm P _ {n+1}}{|\bm P _ n||\bm P _ {n+1}|}
\end{align}
(since  $\bm P_n \times \bm P _ {n+1}
= \mathbf e_z |\bm P_n||\bm P _ {n+1}| \sin \phi$).

Assuming that the wavefunction's weight on each site is approximately the same, we put
\begin{align}
    |\bm{P} _ n|^2 \approx \frac{1}{N}
\end{align}
and hence
\begin{align}
    \frac{1}{\tilde{{P}}} \approx \frac{N}{N-1} \sum _ {n=1} ^{N-1} \bm P _ n \times \bm P _ {n+1}
    \approx \sum _ {n=1} ^ {N-1} \bm P_n \times \bm P_{n+1}
\end{align}
for large $N$. The right hand side, by definition, is the expectation value of the
helicality operator. The left side is the average sine of the angle of precession
per bond.

\subsection{Helicality of cumulene's eigenstates \label{app:previousWork}}
Analytic expression for $\pi$ orbitals of cumulenes were given in
Ref.~\cite{GunasekaranHelical}. In our notation, they are
eigenstates of \pref{eq:fullh} with end-group terms \pref{eq:hend} and $\epsilon\to\infty$.
These eigenstates can be written in the form
\begin{align}
\label{eq:psim}
    \psi _ m = \bm R(\frac\thetat 2)\cdot\begin{pmatrix}
        \cos \phi \cos (k m + \delta) \\
        \sin \phi \sin (k m + \delta) \\
    \end{pmatrix}r \: ,
\end{align}
for the sites $1\le m\le N-1$ (\ie{} excluding end-groups).
In the above equation, the two components denote
the p$_x$ and p$_y$ projections; in our notation,
$\psi_m \equiv \bm P_m$. The $r$ is a normalization factor,
the $\phi$ is the so-called ellipticity angle, $\delta$ is the initial
phase of the molecular orbital, $k$ is its wavenumber. It can be seen
that the wavefunction represents an ellipsis in a two-dimensional space.
Helicality is then associated there with non-trivial values of $\phi \neq \pm \frac{\pi}{2}, 0$.

In the spirit of Sec.~\ref{sec:hop},
we evaluate the axial component of $\bm P_m\times \bm P_{m+1}$, \epref{eq:x}.
The global rotation $\bm R(\frac\thetat 2)$ does not affect the result, which
simplifies to
\begin{align}
    \tilde h_m
    &= r ^ 2 \sin \phi \cos \phi \Bigl\{
        \cos (k m + \delta) \left [ \sin (k m + \delta ) \cos (k) + \right . \nonumber \\
        &+ \left . \left . 
        \cos (k m + \delta ) \sin (k) \right ] - \right . \nonumber \\
        &- \left . \sin(k m + \delta) \left [
            \cos (k m + \delta) \cos (k) \right . \right . \nonumber \\
            &-  \left . \sin (k m + \delta) \sin (k)
        \right ]
    \Bigr \} = \frac{1}{2} r ^ 2 \sin (k) \sin(2 \phi)
\end{align}
Since $0 \le k\le \frac\pi 2$,
the sign of $\tilde h_m$ equals the sign of
$\sin(2 \phi)$. The sign of $\phi$ described the sense of
winding in Gunasekaran \ea{}. Concluding, our operator yields
the winding sense consistently with the latter reference.
Importantly, helicality adopts non-zero expectation values
except for $\phi = \pm \frac{\pi}{2}, 0$.

Simplification occurs for $\phi = \mp \frac \pi 4$, when
vectors $\bm P_m$ draw a circle clockwise 
or anti-clockwise. Using \epref{eq:psim} directly
and estimating $r^2 \approx 2/N$ (from normalisation of eigenstates of $\hat H_0$ \eqref{eq:unperturbedEigenvector}),
we have 
\[ \tilde h_m \approx \mp\frac 1{N} \sin(k). \]
The expectation value of $\hat h$ is obtained by summing
over bonds, which yields
\[ \expval{\hat h} \approx \mp\frac {N-1}{N} \sin(k). \]
For large $N$, we get an average
bond twisting angle $\mp k$.

\subsection{\label{app:averagePitchDensityCommutator}
Commutator of helicality with $\hat H_0$}

The commutator of the helicality operator
$\hat h = \frac{1}{2} \left ( \Tmat - \Tmat ^ \top \right ) \otimes \ci \s{2}$
with $\hat H _ 0$ can be expanded as follows
\begin{align}
    [\hat H_0, \hat h] &= -t\left ( (\Tmat + \Tmat^ \top) \cdot
     (\Tmat -\Tmat ^ \top) \right ) \otimes  \ci \s{0}\s{2} - \nonumber \\
    &- t \left ( (\Tmat -\Tmat ^ \top)\cdot(\Tmat + \Tmat^ \top )\right )
    \otimes \ci \s{2} \s{0} = \nonumber \\
    &= -t \left [\Tmat + \Tmat ^ \top, \Tmat - \Tmat ^ \top \right ]
    \otimes \ci \s{2}.
\end{align}
With the matrix elements
\begin{align}
    (\Tmat) _ {jk} &= \delta _ {j,k-1} \nonumber \\
    (\Tmat ^ \top) _ {jk} &= \delta _ {j-1,k}
\end{align}
we obtain
\begin{align*}
    \left [ \Tmat + \Tmat ^ \top ,\Tmat - \Tmat ^ \top\right ] _ {jk} &=
    \sum _ {l = 1} ^ N (\delta _ {j, l-1} + \delta _ {j - 1, l})(\delta _ {l, k-1} - \delta _ {l-1,k}) \nonumber \\
    &- (\delta _ {j, l-1} - \delta _ {j-1,l})(\delta _ {l, k-1} + \delta _ {l-1,k}) = \nonumber \\
    &= \sum _ {l = 1} ^ N 2 \delta _ {j-1,l} \delta _ {l, k-1} - 2 \delta _ {j, l-1} \delta _ {l-1,k} = \nonumber \\
    &= 2 \sum _ {l = 1} ^ N \delta _ {j,l+1} \delta _ {l+1, k} -
    2 \sum _ {l = 0} ^ {N-1} \delta _ {j, l} \delta _ {l, k} = \nonumber \\
    &= 2 \sum _ {l = 2} ^ {N+1} \delta _ {j, l} \delta _ {l, k} -
    2 \sum _ {l = 1} ^ {N-1} \delta _ {j, l} \delta _ {l, k} = \nonumber \\
    &= 2 \left ( \delta _ {j, N+1} \delta _ {N+1,k} + \delta _ {j, N} \delta _ {N, k} - \delta _ {j, 1} \delta _ {1, k} \right )
\end{align*}
and since $j$ and $k$ are never larger than $N$, we have the expression
\begin{align}
\label{eq:comH0h}
    [\hat H _ 0, \hat h] = (-t) (\Bmat _ N - \Bmat _ 1) \otimes \ci \s{2} \: ,
\end{align}
where $(\Bmat_ l) _ {jk} = \delta _ {j l} \delta _ {l k}$.

\subsection{Extension to carbon rings\label{app:pbc}}
Expression \pref{eq:comH0h} suggests the idea that helicality is an integral
of motion when chains (hard-wall boundaries) are closed into rings with
periodic boundary conditions (PBC) (for example, Garner \ea{} investigate helical
orbitals in cyclic systems \cite{Garner2018}).
To validate this idea mathematically, we introduce the ring Hamiltonian
\begin{align}
    \hat{H}_0^\text{PBC} = -t (\Smat + \Smat^\top) \otimes \s{0}
\end{align}
where
\begin{align*}
    (\Smat)_{jk} = \delta_{j,k\, \mathrm{mod}\, N + 1} \: .
\end{align*}
Here $k\,\mathrm{ mod }\,N$ is the remainder of $k$ after integer division by $N$ ($k$ modulo $N$).
$\hat{H}_0^\text{PBC}$ is essentially $\hat{H}_0$ plus additional matrix elements
that allow for hybridization ('hopping') of carbon 1 with carbon $N$. 
The helicality operator needs to be amended as well in order to account
for the twist of the 1--$N$ bond,
\begin{align*}
    \hat{h}^\text{PBC} = \frac{\ci}{2} (\Smat - \Smat^\top) \otimes \s{2} \:.
\end{align*}

The commutator in PBCs becomes
\begin{align*}
    [\hat{H}_0^\text{PBC}, \hat{h}^\text{PBC}] &=
    \frac{-\ci t}{2} [\Smat + \Smat^\top, \Smat - \Smat^\top] \otimes \s{2} \\
    &= \frac{-\ci t}{2} ([\Smat^\top, \Smat] - [\Smat, \Smat^\top]) \otimes \s{2} = \\
    &= \frac{- \ci t}{2} ([\Smat^\top, \Smat] + [\Smat^\top, \Smat]^\top)
    \otimes \s{2} \: .
\end{align*}
So, we only need to evaluate the elements of $[\Smat^\top, \Smat]$
\begin{align*}
    ([\Smat^\top, \Smat])_{jk} &= \sum _ {m = 1}^N \left (
    \delta_{j\mathrm{mod}N+1,m} \delta_{m, k\mathrm{mod}N+1} \right . \\
    & \left . - \delta_{j, m \mathrm{mod}N + 1} \delta_{m \mathrm{mod} N + 1, k} \right ) \: .
\end{align*}
Since both $m$ and $m \mathrm{mod} N + 1$ produce integers 
from $1$ to $N$ for all $m$ from $1$ to $N$, the summation evaluates to
\begin{align*}
    ([\Smat^\top, \Smat])_{jk} = \delta_{j \mathrm{mod} N + 1,k \mathrm{mod} N + 1} - \delta_{j, k} \: .
\end{align*}
Again, since $j$ and $k$ are both within $1$ and $N$, the first Kronecker
delta reduces to $\delta_{jk}$, which leads to
\begin{align}
    ([\Smat^\top,\Smat])_{jk} = 0
\end{align}
and hence
\begin{align}
    [\hat{H}_0^\text{PBC}, \hat h^\text{PBC}] = 0.
\end{align}

\subsection{Helicality eigenstates\label{app:eigenvectors}}

The eigenvectors of $\hat{h}$, determined analytically, read
\begin{align}
    \braket{n}{n_{hl},n_{hsl},m} = Z_{n_ {hsl}}
    \left(\begin{array}{rl}
      \iu^{n_ {hsl}}  & \sin \left ( \frac{n _ {hl} \pi}{N+1} \right ) \\
      \iu^{2n_ {hsl}} & \sin \left ( \frac{2 n _ {hl} \pi}{N+1} \right ) \\
        \vdots & \\
      \iu^{Nn_ {hsl}} & \sin \left ( \frac{N n _ {hl} \pi}{N+1} \right ) \\
    \end{array}\right) 
    \otimes 
    \begin{pmatrix}
        \frac{1}{\sqrt{2}} \\
        \frac{\ci m}{\sqrt{2}} \\
     \end{pmatrix} \: ,
\end{align}
where the quantum numbers take on integer values
$n_{hl} \in \{1, 2, ..., N/2\}$, $m,n_{hsl}\in \{-1,1\}$.
Notice that the commutator $\com{\hat L_z}{\hat h}=0$ is reflected here:
both operators are diagonalized simultaneously; 
$\hbar m$ is actually the angular momentum projection; the $\hat L_z$ eigenstate 
is the rightmost factor in the above expression.

The corresponding eigenvalues of $\hat h$ are
\begin{align}
    h = - n _ {hsl} m \cos \left ( \frac{n _ {hl} \pi}{N+1}\right ).
\end{align}
We observe that both real and imaginary parts of $\braket{n}{n_{hl},n_{hsl},m}$
have form of circularly-polarized states. Namely, the associated $\bm P_n$ moves
on a circle (notice the power of the imaginary i), making a $\pm$90 degrees turn over each bond.
The circle's radius is scaled by the sine factor on each site.



\section{Impact of sub-lattice symmetry-breaking on the wavefunctions \label{app:sublattice}}
We employ a test Hamiltonian of an $N=4$ chain, 
\begin{equation}
\label{eq:htest}
\hat H_\text{test} =  \hat H_0 + \Bmat_1 \otimes\, \qty( \frac{\epsilon}{2} \s{3} + \delta\s{0})
 + \Bmat_4 \otimes \, \qty(\frac{\epsilon}{2} \s{1} + \delta\s{0}),
\end{equation}
where $\hat H_0$ is given in \epref{eq:hamUnperturbed} and the end-group terms
are expressed as in \epref{eq:hamproductform}. \epref{eq:com1} implies that
the terms proportional to $\delta$ break SL symmetry.
It can be shown that the end-group terms are rotated with respect to
each other by $\thetat=\frac\pi 4$, in the spirit of \epref{eq:hend2}.

Let us have the normalized eigenvectors $\ket{i} (i = 1,\ldots 8)$ of 
$\hat H_\text{test}$
ordered with increasing energy.
When $\delta = 0$, $\ket{i}$ and $\ket{8-i}$ are related
by $\ket{i} = \hat Q\ket{8-i}$, up to a phase factor -- possibly a minus sign.
This motivates the
definition of the measure of SL symmetry breaking
\begin{equation}
\Delta_i\, \coloneqq\, \text{min} \Bigl\|\, \ket{i} \pm \hat Q\ket{8-i}\, \Bigr \|
\end{equation}
for a given orbital pair. The min and $\pm$ are due to the phase ambiguity and the delimiters
indicate that vector norm is to be taken.
\begin{figure}[h]
\includegraphics[width=\columnwidth]{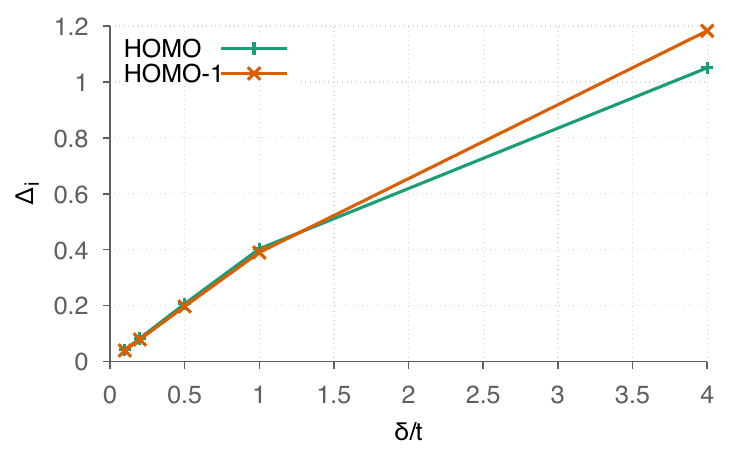}
\caption{\label{fig:subbreak} Norm of the difference of
a Hamiltonian eigenvector, $\ket{i}$, and its SL mirror partner --
see the definition in Appendix~\ref{app:sublattice}. $\delta/t$ controls
the strength of SL breaking terms in the Hamiltonian of
a 4-carbon chain (\epref{eq:htest} with $\epsilon = 0.2$).}
\end{figure}

We show $\Delta_i$ for HOMO and HOMO-1 in \figref{fig:subbreak}.
It grows approximately linearly for small $\delta$ and the
curve is concave, because $\Delta_i$ is bounded from above.
Since the SL breaking terms are local, $\Delta_i$
remains small unless $\delta$ grows beyond $\sim 0.5 t$.
In particular,
for the model of chains with H$_3$C-- end groups, with $\epsilon = 0.2$
and $\delta = \epsilon/2 = 0.1$,
both $\Delta$'s are smaller than 0.1.

\section{Perturbation theory for end-group terms \label{app:perturbation}}
The perturbative treatment of end-groups 
[Eqs.~(\ref{eq:fullh},\ref{eq:hend2},\ref{eq:h1simple})] requires paying special
 attention to the degeneracies; we review the most important formul\ae{} in
 these two subsections.
The reader can find more details in the thesis \cite{MarekThesis}.

\subsection{First order corrections to observables}
Suppose we have an unperturbed Hamiltonian $\hat H ^ 0$
with eigenstates $\ket{\psi ^ 0 _ {\mu l}}$ of $\hat H^0$ coming in degenerate $M-$tuples
\begin{equation}
   \hat H ^ 0 \ket{\psi _ {\mu l}} =
    E ^ 0 _ \mu \ket{\psi _ {\mu l} ^ 0}, \quad l \in \{1, 2, ..., M\},  
\end{equation}
where $\mu$ is the non-degenerate index and $l$ enumerates eigenstates within a  
 degenerate subspace. We assume that a perturbation $\hat V$ fully lifts the degeneracy
of the system. To the first order, an established procedure\cite{AtkinsChemistry,LandauQuantumNonrel}
is to choose a different basis
\begin{equation}
    \ket{\phi ^ 0 _ {\mu j}} = \sum _ {l=1} ^ M \ket{\psi ^ 0 _ {\mu l}} \alpha _ {\mu,lj}
\end{equation}
which diagonalizes the matrix elements of $\hat V$ for each $\mu$, i.e.
\begin{equation}
    \forall \mu: \bra{\phi ^ 0 _ {\mu l}} \hat V \ket{\phi ^ 0 _ {\mu l'}} = \tilde V _ {\mu l,\mu l} \delta _ {ll'} \: ,
\end{equation}
where $\tilde V _ {\mu l,\mu l} = \bra{\phi ^ 0 _ {\mu l}} \hat V \ket{\phi ^ 0 _ {\mu l}}$ and $\delta _ {ll'}$ is the Kronecker delta. In such basis, the perturbation series goes as
\begin{align}
    \left ( \hat H ^ 0 + \hat V \right ) \left ( \ket{\phi ^ 0 _ {\mu l}} + \ket{\phi ^ 1 _ {\mu l}} + ... \right ) = \nonumber \\
    = (E ^ 0 _ \mu + E ^ 1 _ {\mu l} + ...) \left ( \ket{\phi ^ 0 _ {\mu l}} + \ket{\phi ^ 1 _ {\mu l}} + ... \right ) \: .
\end{align}

Subtracting the zeroth order and leaving only terms to first order leads to
\begin{equation}
    \left ( \hat H ^ 0 - E ^ 0 _ \mu \right ) \ket{\phi ^ 1 _ {\mu l}} =
    (E ^ 1 _ {\mu l} -\hat V) \ket{\phi ^ 0 _ {\mu l}}
    \label{firstOrderSchrödinger}
\end{equation}

Scalar product shows that $E ^ 1 _ {\mu l} = \tilde V _ {\mu l,\mu l}$, which is also the eigenvalue of eigenvectors used for diagonalisation of the matrix elements of $\hat V$.

By a general scalar product with $\bra{\phi ^ 0 _ {\mu l'}}$ of the right-hand side of \eqref{firstOrderSchrödinger}, we find that the resulting vector has no contribution in the degenerate space
\begin{equation}
    \bra{\phi ^ 0 _ {\mu l'}} (E ^ 1 _ {\mu l} - \hat V) \ket{\phi ^ 0 _ {\mu l}} = 
    E ^ 1 _ {\mu l} \delta _ {ll'} - \tilde V _ {\mu l',\mu l} = 0
\end{equation}
and therefore we can retrieve the first order corrections to the state as
\begin{equation}
    \ket{\phi ^ 1 _ {\mu l}} = \sum _ {\nu \neq \mu} \sum _ {l' = 1} ^ M \frac{(-\tilde V_{\nu l',\mu l})}{E ^ 0 _ \nu - E ^ 0 _ \mu} \ket{\phi ^ 0 _ {\nu l'}} \: .
\end{equation}
This is a standard result of the perturbation theory, but we explicitly separated the sum over the (originally) degenerate subspaces.

The corrections to observables (such as helicality $\hat h$) are given as
\begin{equation}
    h ^ 0 _ {\mu l} + h ^ 1 _ {\mu l} + ... = \left ( \bra{\phi ^ 0 _ {\mu l}} + \bra{\phi ^ 1 _ {\mu l}} + ...\right )
    \hat h \left ( \ket{\phi ^ 0 _ {\mu l}} + \ket{\phi ^ 1 _ {\mu l}} + ...\right )
\end{equation}
where the zeroth order $h ^ 0 _ {\mu l} = \bra{\phi ^ 0 _ {\mu l}} \hat h \ket{\phi ^ 0 _ {\mu l}}$.

To the first order, we then have (similar as in the non-degenerate case in \cite{LandauQuantumNonrel})
\begin{align}
    h ^ 1 _ {\mu l} &= \bra{\phi ^ 1 _{\mu l}} \hat h \ket{\phi ^ 0 _ {\mu l}} + 
    \bra{\phi ^ 0 _ {\mu l}} \hat h \ket{\phi ^ 1 _{\mu l}} = \\
    &= \sum _ {\nu \neq \mu } \sum _ {l' = 1} ^ M \frac{(-\tilde V_{\nu l',\mu l}) ^ * h _ {\nu l',\mu l} + 
    (-\tilde V _ {\nu l',\mu l}) h _ {\mu l,\nu l'}}{E ^ 0 _ \nu - E ^ 0 _ \mu} \: .
    \label{firstOrderCorrections}
\end{align}

\subsection{Application to oligoynes with methyl end-groups\label{app:perturbationApplication}}
The Hamiltonian $\hat H_0$ has eigenstates of the form (for even $N$) \cite{GunasekaranHelical}
\begin{equation}
    \ket{\psi _ {\mu l} ^ 0} = Z \begin{pmatrix}
        \sin \left( \frac{\mu \pi}{N+1} \right ) \\
        \sin \left ( \frac{2 \mu \pi}{N+1} \right ) \\
        \vdots \\
        \sin \left ( \frac{N \mu \pi}{N+1} \right )  \\
    \end{pmatrix}  \otimes \bm v _ l, \quad {{1\le \mu \le N,}\atop {l=\pm 1}},
    \label{eq:unperturbedEigenvector}
\end{equation}
where $Z$ is the normalization constant $Z = \sqrt{2/(N+1)}$ and $\bm v _ l$ is a 2 component vector. The eigenstates are degenerate in components of $\bm v _ l$, hence we can choose 2 orthogonal vectors $\bm v _ \alpha$ and $\bm v _ \beta$ to get an orthogonal degenerate eigenstates.
The energies of the eigenstates $\ket{\psi^0_{\mu l}}$ are (independent of $l$)
\begin{align}
	E^0_{\mu l} = - 2 t \cos \left ( \frac{\pi \mu}{N + 1} \right ) \: .
\end{align}

Now we choose the Hamiltonian \pref{eq:fullh} with the perturbation $\hat H_1 +\hat H_N$,
parameterized by \epref{eq:hend2} (Model 2).
Solving the eigenvalue equation to diagonalize matrix elements of $\hat H _ \text{pert}$ leads to eigenvectors of form
\begin{equation}
    \bm v _ \alpha = \begin{pmatrix}
        \sin\left (\theta_\mathrm{T}/2\right ) \\
        -\cos \left ( \theta_\mathrm{T}/2 \right )\\
    \end{pmatrix} \: , \:
    \bm v _ \beta = \begin{pmatrix}
        \cos \left ( \theta_\mathrm{T}/2 \right )\\
        \sin\left ( \theta_\mathrm{T}/2 \right ) \\
    \end{pmatrix}
\end{equation}
which leads to eigenvalues mentioned in main text
\begin{equation}
    E ^ 1 _ {\mu, \alpha/\beta} = 2 \epsilon Z ^ 2 \sin ^ 2 \left ( \frac{\mu \pi}{N+1} \right )
    (\cos / \sin) ^ 2 \left ( \theta_\mathrm{T}/2 \right ) \: .
\end{equation}
\begin{figure}[ht]
    \centering
    \includegraphics[width=\columnwidth]{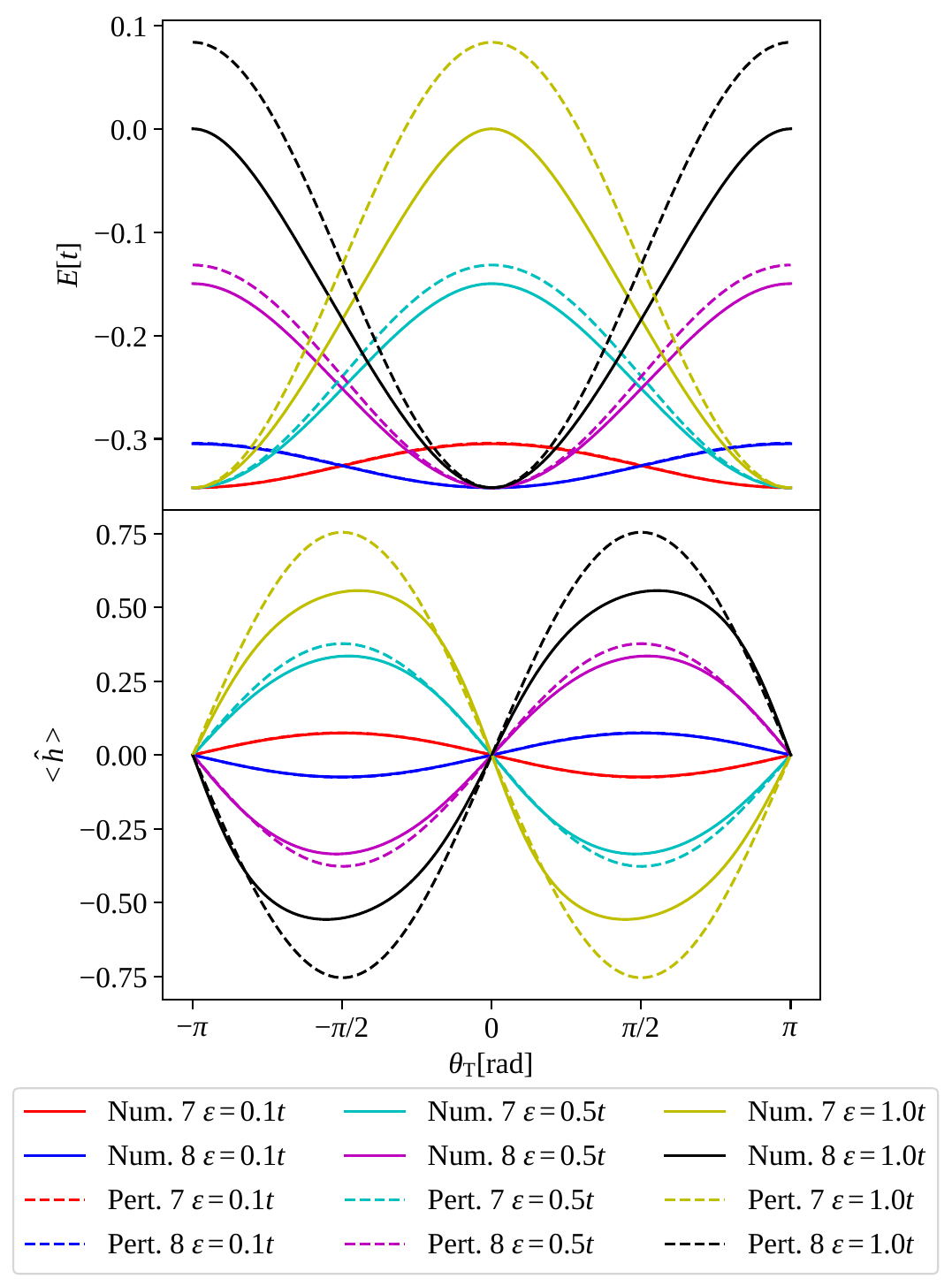}
    \caption{The energy and helicality $\expval{\hat h}$ of a doublet of eigenstates, HOMO and HOMO-1,
     calculated by first-order perturbation theory and numerically by exact diagonalization, for $N=8$.
     The end-group term with $\epsilon = 0.15t$ and varying twist angle $\thetat$ is added.
  }
    \label{perturbationGraph}
\end{figure}

Hence, the natural eigenstates for the perturbation are two perpendicular 
"linearly polarized"\cite{GunasekaranHelical} strands of p-orbitals. 
The first order corrections to helicality calculated by \eqref{firstOrderCorrections}
are then readily obtained. The results are presented graphically for HOMO and 
HOMO-1 in \figref{perturbationGraph}.

\FloatBarrier

\begin{figure}[h]
    \centering
    \includegraphics[width=\columnwidth]{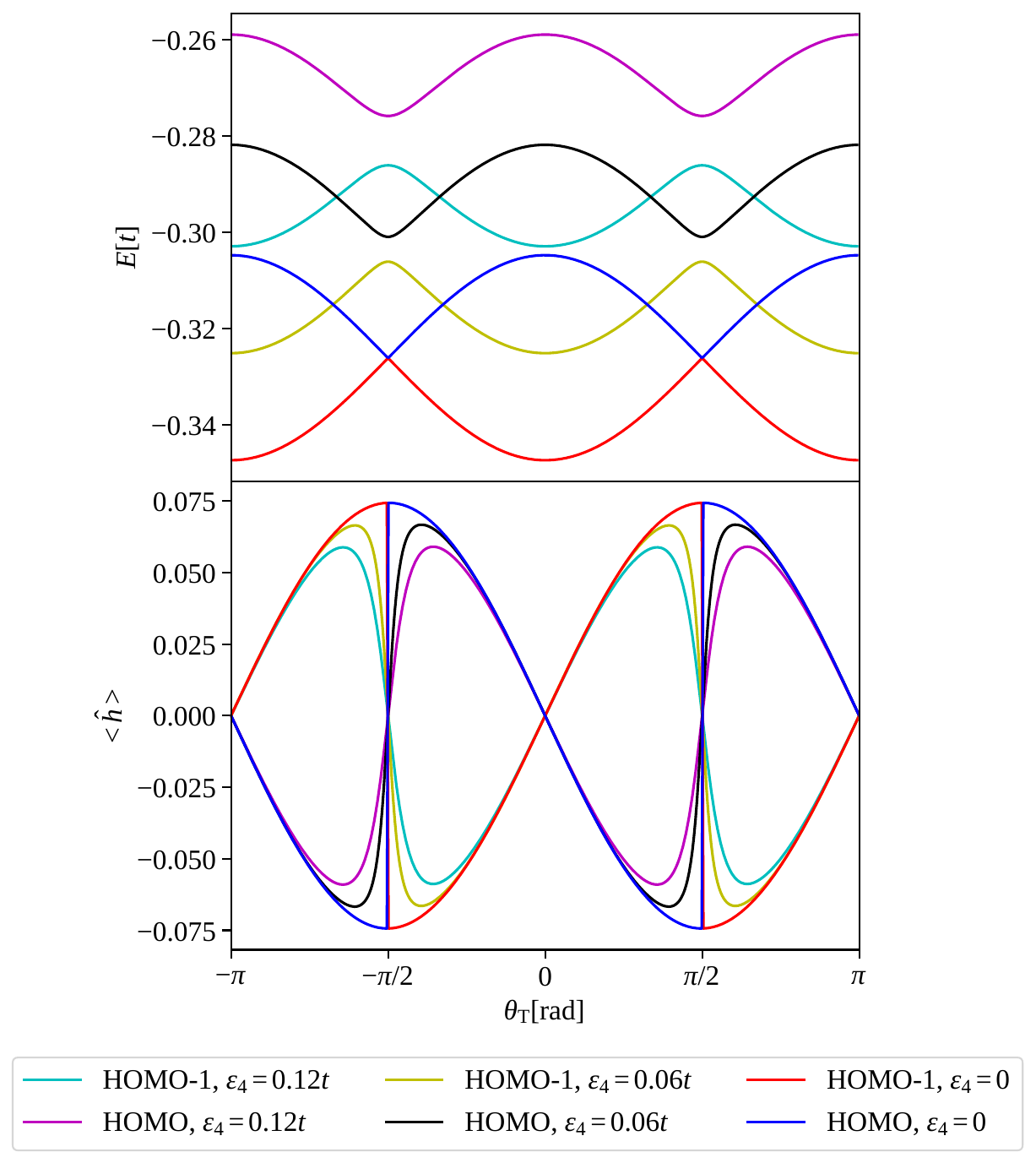}
    \caption{\revision{}{The energy (top) and the 
    expectation value of helicality (bottom) for several
	values of the onsite potential. Compared to the ordered case ($\epsilon_4 = 0$), the
    gap between the eigenstates opens in place of the crossings. The helicality of the pair
    of states retains opposite sign at any chosen angle $\theta_\mathrm{T}$. Here, number of sites
    $N=8$ and $\epsilon = 0.1 t$.}}
    \label{fig:onsite}
\end{figure}

\begin{figure}[h]
    \centering
    \includegraphics[width=\columnwidth]{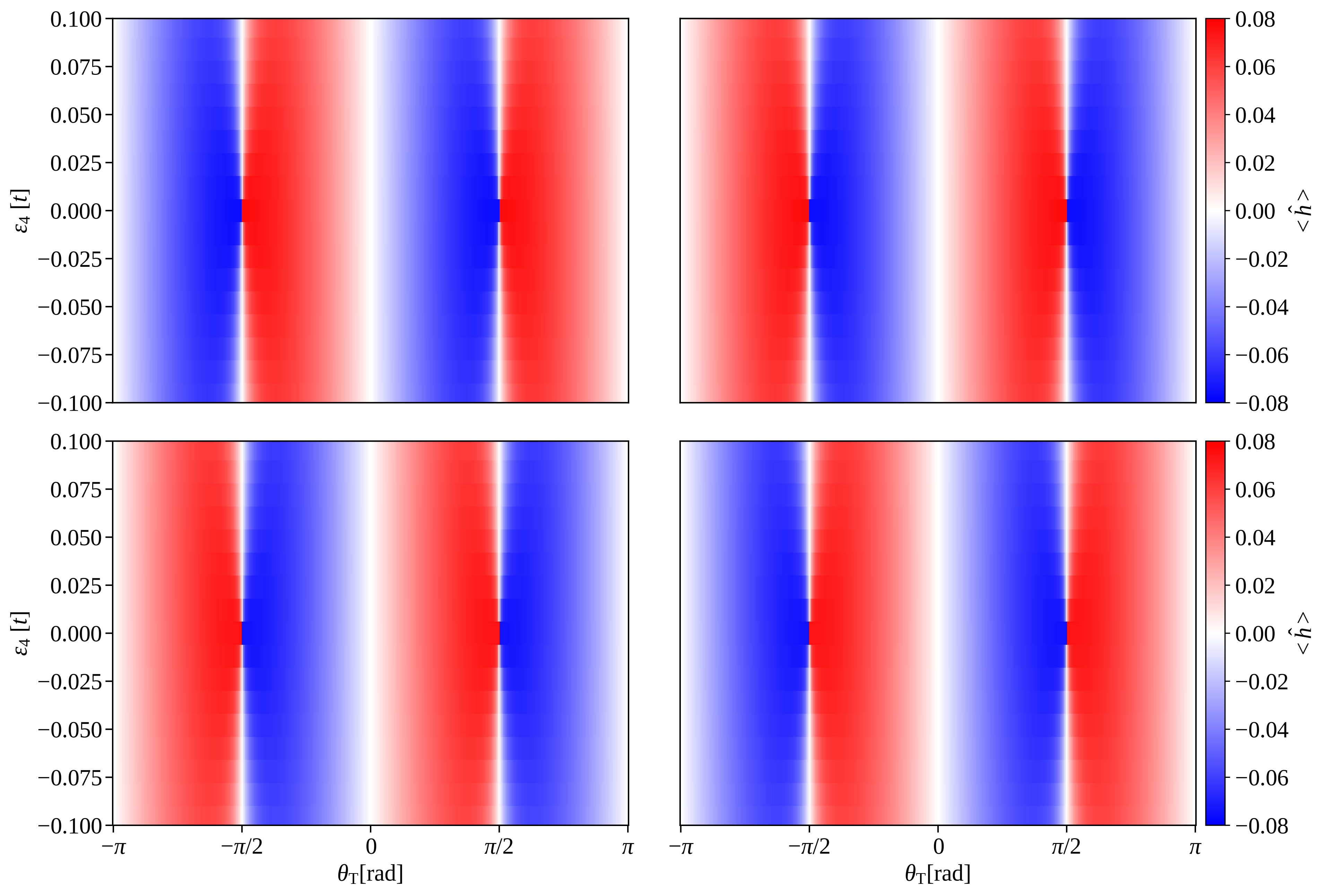}
    \caption{\revision{}{The expectation value of helicality for HOMO-1,HOMO (bottom row) and
    LUMO+1,LUMO in a chain of length $N=8$ with $\epsilon = 0.1 t$. We observe that
    the opposite sign of helicality between the orbital pairs is present.}
    \label{fig:onsiteMap}}
\end{figure}

\revision{}{\section{\label{app:dis}Substitutional disorder}

One way to validate robustness of our results is to investigate presence of onsite substitution
in the chain. We model the substitution only in a very simplified way -- 
as a presence of an onsite potential on a single atom in the chain.

We chose the chain of length $N=8$, as in the previous section, with $\epsilon = 0.1 t$
and swept through all $\theta_\mathrm{T}$ while also varying the onsite potential $\epsilon_4$.
The Hamiltonian reads
\begin{align}
    \hat H \to \hat H + \epsilon _ 4 \mathbb B_4 \otimes \hat{\sigma}_0 \: .
\end{align}
The resulting energy of selected eigenstates and the expectation value of helicality in these
eigenstates is presented in \figref{fig:onsite}.
We notice the avoided crossings at $90^\circ$. 
Results for broader choices of $\epsilon_4$ are shown in \figref{fig:onsiteMap}. Again, the opposite signs
of helicality in orbital pairs is observed.}

\revision{}{\section{\label{app:vibrations} Effect of molecular vibrations}
The Onsager relations derived in this work rely on time-reversal
invariance and sub-lattice (SL) symmetry. Coupling of electrons
to molecular vibrations preserves the former but the fate of the
latter is discussed here briefly.

To investigate the effects of electron-vibrational coupling, we
reformulate the Hamiltonian from first to second quantization, namely,
\begin{equation*}
\hat{\mathcal H} = \sum_{I,I'} H_{II'}
\cc_I \ca_{I'},
\end{equation*}
where $\cc_I$ and $\ca_I$ denote fermionic creation and annihilation operator;  the single-particle basis is indexed by $\ket{I} \equiv \ket{\alpha n}$ (see above \epref{eq:hamUnperturbed}),
and $H_{II'} = \bra{I}\hat H\ket{I'}$.
The sub-lattice (SL) symmetry is a condition expressed on the 'first-quantized'
matrices as $\sum_{I} \qty( H_{I'I}P_{II''} + H_{II''}P_{I'I})=0$ with 
$P_{II'} = \bra{I}\hat P\ket{I'}$
as given in \eqref{eq:pdef} (alternatively, \eqref{eq:qdef}). 
The Hamiltonian that describes
molecular vibrations, the electronic system and  their linearized
mutual coupling takes generally the form
\begin{equation}
\hat {\mathcal K} = \hat {\mathcal H}\, +\, \hat {\mathcal H}_\text{vib}\qty(\ba\!,\,\bc)
\,+\, \sum_{II'} M^v_{II'} \cc_I \ca_{I'} \qty(\ba_v + \bc_v),
\end{equation}
where $\bc_v$ and $\ba_v$ denote bosonic creation and annihilation operators. The electron-vibrational
coupling preserves SL symmetry when the coupling elements obey
\begin{equation}
\label{eq:MP}
\sum_{I} \qty( M^v_{I'I}P_{II''} + M^v_{II''}P_{I'I})=0,
\end{equation}
\ie{}, when the corresponding matrices anti-commute. 
The deviation from the above rule occurs \eg{} when the vibration 
induces a strong inhomogeneity implying that $M^v_{II}$ is different for different sites.

The interaction of electrons and vibrations in polyynes (infinitely long carbon wires with alternating single and triple bonds)
was investigated by Rice \ea{} \cite{Rice1986}. The authors used a
Su-Schrieffer-Heeger \cite{Heeger1988} (SSH) model with two electronic bands,
p$_x$ and p$_y$ and a coupling to longitudinal vibrations incorporated as a modulation of the nearest-neighbor
hopping elements. 
The interaction with transverse vibrations is often considered
much weaker, as transverse vibrations have a smaller impact on
the effective electrostatic potential \cite{Kittel1963}.
Crucially, the modulation of nearest-neighbor hopping amplitudes satisfies
 condition \pref{eq:MP}. An analogous situation occurs in graphene, where
 the dominant electron-vibrational coupling was assumed to be SL preserving \cite{Ando2008,Mariani2008}.
 We conclude that SL symmetry will be preserved
in vibrating chains as long as the SSH description is valid, even if the lattice
deformation is large and the coupling is strong. In this regime,
the Onsager relations are expected to hold since the underlying microscopic symmetries are preserved.
}

\FloatBarrier

\bibliography{sources}
\end{document}